\documentclass[twocolumn]{aastex631}

\usepackage{graphicx}	
\usepackage{amssymb}	
\usepackage{xcolor}
\usepackage{subfigure}
\usepackage{verbatim}
\usepackage{amsmath}
\usepackage{float}
\usepackage{soul}
\usepackage{gensymb}
\usepackage{amsmath}
\usepackage{mathtools} 
\usepackage[normalem]{ulem}
\usepackage{soul}
\usepackage{amssymb}

\newcommand*{\rom}[1]{\expandafter\@slowromancap\romannumeral #1@}
\newcommand{\cmmnt}[1]{}

\begin{document}
\title[AGN STORM 2]{AGN STORM 2: IX. Studying the Dynamics of the Ionized Obscurer in Mrk~817 with High-resolution X-ray Spectroscopy}


\author[0000-0003-0931-0868]{Fatima Zaidouni}
\affiliation{MIT Kavli Institute for Astrophysics and Space Research, Massachusetts Institute of Technology, Cambridge, MA 02139, USA}

\author[0000-0003-0172-0854]{Erin Kara}
\affiliation{MIT Kavli Institute for Astrophysics and Space Research, Massachusetts Institute of Technology, Cambridge, MA 02139, USA}

\author[0000-0003-4511-8427]{Peter Kosec}
\affiliation{MIT Kavli Institute for Astrophysics and Space Research, Massachusetts Institute of Technology, Cambridge, MA 02139, USA}

\author[0000-0002-4994-4664]{Missagh Mehdipour}
\affiliation{Space Telescope Science Institute, 3700 San Martin Drive, Baltimore, MD 21218, USA}

\author[0000-0001-9735-4873]{Daniele Rogantini}
\affiliation{MIT Kavli Institute for Astrophysics and Space Research, Massachusetts Institute of Technology, Cambridge, MA 02139, USA}

\author[0000-0002-2180-8266]{Gerard A.\ Kriss}
\affiliation{Space Telescope Science Institute, 3700 San Martin Drive, Baltimore, MD 21218, USA}

\author[0000-0001-9735-4873]{Ehud Behar}
\affiliation{Department of Physics, Technion, Haifa 32000, Israel}
\affiliation{MIT Kavli Institute for Astrophysics and Space Research, Massachusetts Institute of Technology, Cambridge, MA 02139, USA}

\author[0000-0001-5540-2822]{Jelle Kaastra}
\affiliation{SRON Netherlands Institute for Space Research, Niels Bohrweg 4, 2333 CA Leiden, The Netherlands}
\affiliation{Leiden Observatory, Leiden University, PO Box 9513, 2300 RA Leiden, The Netherlands}

\author[0000-0002-3026-0562]{Aaron J.\ Barth}
\affiliation{Department of Physics and Astronomy, 4129 Frederick Reines Hall, University of California, Irvine, CA, 92697-4575, USA}

\author[0000-0002-8294-9281]{Edward M.\ Cackett}
\affiliation{Department of Physics and Astronomy, Wayne State University, 666 W.\ Hancock St, Detroit, MI, 48201, USA}

\author[0000-0003-3242-7052]{Gisella De~Rosa}
\affiliation{Space Telescope Science Institute, 3700 San Martin Drive, Baltimore, MD 21218, USA}

\author[0000-0002-0957-7151]{Yasaman Homayouni}
\affiliation{Space Telescope Science Institute, 3700 San Martin Drive, Baltimore, MD 21218, USA}
\affiliation{Department of Astronomy and Astrophysics, The Pennsylvania State University, 525 Davey Laboratory, University Park, PA 16802}
\affiliation{Institute for Gravitation and the Cosmos, The Pennsylvania State University, University Park, PA 16802}

\author[0000-0003-1728-0304]{Keith Horne}
\affiliation{SUPA School of Physics and Astronomy, North Haugh, St.~Andrews, KY16~9SS, Scotland, UK}

\author{Hermine Landt}
\affiliation{Centre for Extragalactic Astronomy, Department of Physics, Durham University, South Road, Durham DH1 3LE, UK}

\author[0000-0003-2991-4618]{Nahum Arav}
\affiliation{Department of Physics, Virginia Tech, Blacksburg, VA 24061, USA}


\author[0000-0002-2816-5398]{Misty C.\ Bentz}
\affiliation{Department of Physics and Astronomy, Georgia State University, 25 Park Place, Suite 605, Atlanta, GA 30303, USA}


\author[0000-0002-1207-0909]{Michael S.\ Brotherton}
\affiliation{Department of Physics and Astronomy, University of Wyoming, Laramie, WY 82071, USA}


\author[0000-0001-9931-8681]{Elena Dalla Bont\`{a}}
\affiliation{Dipartimento di Fisica e Astronomia ``G.\  Galilei,'' Universit\`{a} di Padova, Vicolo dell'Osservatorio 3, I-35122 Padova, Italy}
\affiliation{INAF-Osservatorio Astronomico di Padova, Vicolo dell'Osservatorio 5 I-35122, Padova, Italy}

\author[0000-0002-0964-7500]{Maryam Dehghanian}
\affiliation{Department of Physics, Virginia Tech, Blacksburg, VA 24061, USA}

\author[0000-0003-4503-6333]{Gary J.\ Ferland}
\affiliation{Department of Physics and Astronomy, The University of Kentucky, Lexington, KY 40506, USA}


\author[0000-0002-2306-9372]{Carina Fian}
\affiliation{Departamento de Astronom\'{i}a y Astrof\'{i}sica, Universidad de Valencia, E-46100 Burjassot, Valencia, Spain}
\affiliation{ Observatorio Astron\'{o}mico, Universidad de Valencia, E-46980 Paterna, Valencia, Spain}



\author[0000-0001-9092-8619]{Jonathan Gelbord}
\affiliation{Spectral Sciences Inc., 4 Fourth Ave., Burlington, MA 01803, USA}

\author[0000-0002-2908-7360]{Michael R.\ Goad}
\affiliation{School of Physics and Astronomy, University of Leicester, University Road, Leicester, LE1 7RH, UK}

\author[0000-0002-9280-1184]{Diego H.\ Gonz\'{a}lez Buitrago}
\affiliation{Instituto de Astronom\'{\i}a, Universidad Nacional Aut\'{o}noma de M\'{e}xico, Km 103 Carretera Tijuana-Ensenada, 22860 Ensenada B.C., M\'{e}xico}

\author[0000-0001-9920-6057]{Catherine J. Grier}
\affiliation{Department of Astronomy, University of Wisconsin-Madison, Madison, WI 53706, USA}

\author[0000-0002-1763-5825]{Patrick B.\ Hall}
\affiliation{Department of Physics and Astronomy, York University, Toronto, ON M3J 1P3, Canada}


\author{Chen Hu}
\affiliation{Key Laboratory for Particle Astrophysics, Institute of High Energy Physics, Chinese Academy of Sciences, 19B Yuquan Road,\\ Beijing 100049, People's Republic of China}

\author[0000-0002-1134-4015]{Dragana Ili\'{c}}
\affiliation{University of Belgrade - Faculty of Mathematics, Department of astronomy, Studentski trg 16, 11000 Belgrade, Serbia}
\affiliation{Humboldt Research Fellow, Hamburger Sternwarte, Universit{\"a}t Hamburg, Gojenbergsweg 112, 21029 Hamburg, Germany}


\author[0000-0002-9925-534X]{Shai Kaspi}
\affiliation{School of Physics and Astronomy and Wise observatory, Tel Aviv University, Tel Aviv 6997801, Israel}

\author[0000-0001-6017-2961]{Christopher S.\ Kochanek}
\affiliation{Department of Astronomy, The Ohio State University, 140 W.\ 18th Ave., Columbus, OH 43210, USA}
\affiliation{Center for Cosmology and AstroParticle Physics, The Ohio State University, 191 West Woodruff Ave., Columbus, OH 43210, USA}


\author[0000-0001-5139-1978]{Andjelka B. Kova{\v c}evi{\'c}}
\affiliation{University of Belgrade - Faculty of Mathematics, Department of astronomy, Studentski trg 16, 11000 Belgrade, Serbia}

\author[0000-0001-8638-3687]{Daniel Kynoch}
\affiliation{School of Physics and Astronomy, University of Southampton, Highfield, Southampton SO17 1BJ, UK}
\affiliation{Astronomical Institute of the Czech Academy of Sciences, Boční II 1401/1a, CZ-14100 Prague, Czechia}


\author[0000-0002-8671-1190]{Collin Lewin}
\affiliation{MIT Kavli Institute for Astrophysics and Space Research, Massachusetts Institute of Technology, Cambridge, MA 02139, USA}






\author[0000-0001-5639-5484]{John Montano}
\affiliation{Department of Physics and Astronomy, 4129 Frederick Reines Hall, University of California, Irvine, CA, 92697-4575, USA}

\author[0000-0002-6766-0260]{Hagai Netzer}
\affiliation{School of Physics and Astronomy and Wise observatory, Tel Aviv University, Tel Aviv 6997801, Israel}

\author[0000-0001-7351-2531]{Jack M. M. Neustadt}
\affiliation{Department of Astronomy, The Ohio State University, 140 W. 18th Ave., Columbus, OH 43210, USA}


\author{Christos Panagiotou}
\affiliation{MIT Kavli Institute for Astrophysics and Space Research, Massachusetts Institute of Technology, Cambridge, MA 02139, USA}

\author[0000-0003-1183-1574]{Ethan R. Partington}
\affiliation{Department of Physics and Astronomy, Wayne State University, 666 W.\ Hancock St, Detroit, MI, 48201, USA}

\author[0000-0002-2509-3878]{Rachel Plesha}
\affiliation{Space Telescope Science Institute, 3700 San Martin Drive, Baltimore, MD 21218, USA}


\author[0000-0003-2398-7664]{Luka \v{C}.\ Popovi\'{c}}
\affiliation{Astronomical Observatory, Volgina 7, 11060 Belgrade, Serbia}
\affiliation{University of Belgrade - Faculty of Mathematics, Department of astronomy, Studentski trg 16, 11000 Belgrade, Serbia}

\author[0000-0002-6336-5125]{Daniel Proga}
\affiliation{Department of Physics \& Astronomy, 
University of Nevada, Las Vegas 
4505 S.\ Maryland Pkwy, 
Las Vegas, NV, 89154-4002, USA}



\author[0000-0003-1772-0023]{Thaisa Storchi-Bergmann}
\affiliation{Departamento de Astronomia - IF, Universidade Federal do Rio Grande do Sul, CP 150501, 91501-970 Porto Alegre, RS, Brazil}

\author[0000-0002-9238-9521]{David Sanmartim}
\affiliation{Rubin Observatory Project Office, 950 N. Cherry Ave., Tucson, AZ 85719, USA} 

\author[0000-0003-2445-3891]{Matthew R. Siebert}
\affiliation{Department of Astronomy and Astrophysics, University of California, Santa Cruz, CA 92064, USA}

\author[0000-0002-8177-6905]{Matilde Signorini}
\affiliation{Dipartimento di Fisica e Astronomia, Università di Firenze, via G. Sansone 1, 50019 Sesto Fiorentino, Firenze, Italy}
\affiliation{INAF - Osservatorio Astrofisico di Arcetri, Largo Enrico Fermi 5, I-50125 Firenze, Italy}
\affiliation{University of California-Los Angeles, Department of Physics and Astronomy, PAB, 430 Portola Plaza, Box 951547, Los Angeles, CA 90095-1547, USA}



\author[0000-0001-9191-9837]{Marianne Vestergaard}
\affiliation{Steward Observatory, University of Arizona, 933 North Cherry Avenue, Tucson, AZ 85721, USA}
\affiliation{DARK, The Niels Bohr Institute, University of Copenhagen, Jagtvej 155, DK-2200 Copenhagen, Denmark}



\author[0000-0002-5205-9472]{Tim Waters}
\affiliation{Department of Physics \& Astronomy, 
University of Nevada, Las Vegas 
4505 S. Maryland Pkwy, 
Las Vegas, NV, 89154-4002, USA}






\author[0000-0001-6966-6925]{Ying Zu}
\affiliation{Department of Astronomy, School of Physics and Astronomy, Shanghai Jiao Tong University, Shanghai 200240, China}


\begin{abstract}

We present the results of the XMM-Newton and NuSTAR observations taken as part of the ongoing, intensive multi-wavelength monitoring program of the Seyfert 1 galaxy Mrk 817 by the AGN Space Telescope and Optical Reverberation Mapping 2 (AGN STORM 2) Project. The campaign revealed an unexpected and transient obscuring outflow, never before seen in this source. Of our four XMM-Newton/NuSTAR epochs, one fortuitously taken during a bright X-ray state has strong narrow absorption lines in the high-resolution grating spectra. From these absorption features, we determine that the obscurer is in fact a multi-phase ionized wind with an outflow velocity of $\sim$5200 km s$^{-1}$, and for the first time find evidence for a lower ionization component with the same velocity observed in absorption features in the contemporaneous HST spectra. This indicates that the UV absorption troughs may be due to dense clumps embedded in diffuse, higher ionization gas responsible for the X-ray absorption lines of the same velocity. \textcolor{black}{We observe variability in the shape of the absorption lines on timescales of hours, placing the variable component at roughly 1000 $R_g$ if attributed to transverse motion along the line of sight. This estimate aligns with independent UV measurements of the distance to the obscurer suggesting an accretion disk wind at the inner broad line region. We estimate that it takes roughly 200 days for the outflow to travel from the disk to our line of sight, consistent with the timescale of the outflow’s column density variations throughout the campaign.}

\end{abstract}


\let\clearpage\relax
\section{Introduction}
Astrophysical black holes address many significant research questions from tests of theories of gravity (e.g, \citealt{2016CQGra..33e4001Y,2017ApJ...844L..14S}), to investigations of extreme physical processes, such as relativistic jets and outflows (\citealt{2019ARA&A..57..467B,2021NatAs...5...13L}), and their effect on the formation and evolution of galaxies (e.g., \citealt{2009Natur.460..213C}). The latter is due to interactions between the central supermassive black hole (SMBH) and its host galaxy. This phenomenon, referred to as Active Galactic Nucleus (AGN) feedback (e.g, \citealt{2012ARA&A..50..455F}), encompasses the influence of the AGN activity, driven by the SMBH, on the surrounding galactic environment.

AGN feedback helps explain numerous observational results such as the black hole M-$\sigma$ relation and the luminosity function of massive galaxies. The black hole M~-~$\sigma$ relation indicates that the velocity dispersion of the stars in the galaxy bulge is correlated with the mass of the central supermassive black hole (e.g, \citealt{2013ARA&A..51..511K}). The galaxy luminosity function is found to diverge from the Cold Dark Matter (CDM) prediction at high masses (\citealt{1999ApJ...522...82K}). This suggests that massive and powerful outflows from the central black hole help prevent gas in the galaxy from cooling to the level required for star formation (\citealt{2012ARA&A..50..455F}). 

Despite the importance of AGN feedback in explaining multiple results, observational  evidence for the driving mechanism remains sparse. One of the observed mechanisms affecting galaxy star formation is galactic outflows. These large scale outflows have been detected in some AGN (e.g, \citealt{2010A&A...518L.155F,2011ApJ...733L..16S,2014A&A...562A..21C}). However, it is still unclear how these outflows are connected to the central AGN. 

In theoretical models, outflows can drive feedback on large scales when their kinetic luminosities range from  $0.5\%$ to $5\%$ of the Eddington luminosity of the central black hole
(e.g., \citealt{2005Natur.433..604D,2010MNRAS.401....7H, 2020NatAs...4...10G}). In the sphere of influence of the AGN, several types of outflows are detected using high-resolution UV and X-ray spectroscopy; First, Ultra-fast outflows (UFOs) are relativistic, highly ionized (log $\xi = 3-5$ where  the ionization parameter $\xi$ is in units of erg cm s$^{-1}$) outflows mainly detected in the X-rays through highly blueshifted Fe–K lines (e.g., \citealt{2003ApJ...593L..65R,2013MNRAS.430.1102T,2017Natur.543...83P}). On occasion, UV counterparts have been detected (e.g., \citealt{2018ApJ...853..166K,2022ApJ...930..166M}). UFOs have significant kinetic power, but their persistence beyond the environment of the black hole is uncertain. Second, warm absorbers (e.g., \citealt{2003ARA&A..41..117C,2005A&A...431..111B}) have a broad ionization distribution (log $\xi = -1$--$4$, \citealt{2007ApJ...663..799H}) and velocities of hundreds km~s$^{-1}$. They have been detected in both the X-rays and UV (e.g, \citealt{1997ApJ...478..182M, 2012A&A...542A..30M, 2013MNRAS.435.3028E, 2014MNRAS.441.2613L}). They can potentially carry away a significant fraction of the originally infalling mass but they do not have sufficient kinetic luminosity to drive feedback (e.g., \citealt{2005A&A...431..111B,2010SSRv..157..265C}). Finally, obscuring outflows originate from the accretion disk with velocities of thousands of km~s$^{-1}$ and are observed to be recurrent in a number of AGN (e.g, \citealt{1994ApJ...421...69L, 2014Sci...345...64K, 2003MNRAS.342L..41L, 2016A&A...586A..72E, 2017A&A...607A..28M}). They may be a promising driver of AGN feedback, owing to the large momentum they carry.

The origin and driving mechanism of these AGN outflows are unclear. Proposed mechanisms range from line-driving, where line radiation pressure from the disk and central AGN launches unsteady fast streams (e.g., \citealt{2000ApJ...543..686P,2004ApJ...616..688P})  to magnetic driving such as magnetohydrodynamic wind solutions (e.g., \citealt{1982MNRAS.199..883B,1994ApJ...434..446K,2010ApJ...715..636F}), and dust-driven winds where radiation pressure acts on dust grains within the AGN's circumnuclear region (e.g., FRADO model: \citealt{2011A&A...525L...8C,2015ApJ...806..129E}). Thermal driving, where gas is heated by the central engine causing it to expand and flow outwards, may be a viable mechanism for the slower warm absorbers (\citealt{2021ApJ...914...62W}).

The physical connection between the different types of outflows found in X-rays and their relation to the UV broad-line absorbing outflows is also uncertain. More observations spanning diverse scales, and hence multiple wavelengths are needed to discern the outflow properties and their ability to cause AGN feedback. This was one of the main motivations behind the Space Telescope and Optical Reverberation Mapping Projects (AGN STORM 1 and 2); two large multi-wavelength monitoring campaigns discussed below.

AGN STORM 1 targeted the Seyfert 1 galaxy NGC~5548 ($z= 0.0172$) through a multiwavelength spectroscopic and photometric monitoring campaign (\citealt{2015ApJ...806..128D}). It was rooted around daily Hubble Space Telescope (HST) observations for a period of six months in 2014 along with high cadence Swift observations, Chandra observations, and extensive ground-based spectroscopic and photometric monitoring (\citealt{2015ApJ...806..129E, 2017ApJ...837..131P,2016ApJ...821...56F}).

Among the various results of the AGN STORM 1 campaign were changes in the emission and absorption lines (\citealt{2017ApJ...837..131P, 2016ApJ...824...11G}) that are consistent with a variable, obscuring disk wind, launched from the inner broad line region (BLR). This obscurer was previously detected by \citet{2014Sci...345...64K} in a multiwavelength campaign between 2013 May and 2014 Feb centered on 14 XMM-Newton spectra. They observed a persistent, clumpy, ionized outflow responsible for the broad UV absorption lines and for suppressing most of the soft X-ray emission. This obscurer was presented as a likely explanation for the “BLR holiday” observed during AGN STORM 1, a period where the variability of the broad emission lines and the high ionization absorption lines was decoupled from the variability of the continuum (\citealt{2016ApJ...824...11G, 2019ApJ...877..119D}). 

In light of the wealth of results on the multi-wavelength reverberation mapping campaign of NGC~5548 (AGN STORM 1) and the separate XMM campaign for the same target (\citealt{2014Sci...345...64K}), we aimed to combine these efforts into a large campaign that combined both high cadence, multi-wavelength monitoring with several deep XMM/NuSTAR observations. The result is the AGN STORM 2 campaign on Mrk~817 (PG 1434+590, $z=0.0315$, \citealt{1988AJ.....95.1602S}). Mrk~817 was selected to map the accretion disk of an AGN with a significantly higher Eddington ratio than NGC~5548 through reverberation mapping (see \citet{2021iSci...24j2557C} for a recent review on this technique). Studying an ionized obscurer was not the original motivation for the AGN STORM 2 campaign, as this source historically did not show strong absorption in the UV and X-rays (\citealt{2011ApJ...728...28W}). However, at the start of the campaign in November 2020, the source surprisingly showed significant obscuration (\citealt{2021ApJ...922..151K}:~Paper~1).

AGN STORM 2 is also centered around 2-day cadence Hubble Space Telescope observations as part of a large 198-orbit program. Results from the first third of the campaign were presented in Paper~1. In Paper 2, \citet{2023arXiv230211587H} presented the results of the HST UV broad emission line reverberation mapping program. Paper 3, \citet{2023arXiv230212896P}, presented the results of the high-cadence (1--2 day) NICER X-ray campaign, that tracked the spectral variability of the ionized obscurer. Paper 4, \citet{2023arXiv230617663C}, presents the initial continuum reverberation mapping results from the Swift X-ray, UV and optical campaign, which will be followed up with the ground-based optical and near infrared continuum reverberation (Montano et al., in prep.) and spectroscopic broad line region reverberation mapping (Hue et al., in prep.). In this paper, we present the first results on the deep X-ray campaign, which consisted of four simultaneous XMM-Newton and NuSTAR observations, one of which was concurrent with a Chandra/HETG observation. The focus of this paper is the highest flux observation taken fortuitously during the brightest X-ray state observed during the campaign on 2021~Apr.

Paper~1 shows that Mrk~817 was in a deep X-ray low flux state. Various observational results including the variability of the soft X-rays throughout the campaign relative to archival XMM-Newton observations from 2009 (shown in Fig.~\ref{fig:alldata}), the sudden presence of broad and narrow absorption features in the UV data, and the decoupling of the UV continuum and broad emission line variability all pointed towards significant obscuration of the ionizing source, particularly in the X-rays. This obscuration is consistent with a dust-free, ionized, partially covering obscurer located at the inner broad line region (Paper~1). An independent analysis  of  a  concurrent  NuSTAR observation also observed high obscuration (\citealt{2021ApJ...911L..12M}).

In this paper, we use high resolution X-ray spectroscopy to study the properties of the Mrk~817 obscurer. The HST, XMM-Newton/pn, and NuSTAR observations allow us to constrain the broadband spectral energy distribution. The high resolution data from the XMM-Newton Reflection Grating Spectrometer (RGS) instrument allow us to measure outflow properties such as velocity, ionization, and column density. In Section~\ref{sec:data}, we present the observations included in the AGN STORM 2 X-ray campaign and the data reduction procedure. We present the results of our photoionization modeling analysis of the 2021~Apr observation in Section~\ref{sec:results} which we discuss and interpret in Section~\ref{sec:discussion}. We summarize our findings in Section 5.

\begin{table}
\caption{Details of the XMM-Newton PN observations.}\label{tab:observations}
\centering
\begin{tabular}{cccc}
\hline \hline ObsID & Start Time & Net Exposure (ks) \\
\hline  
0872390901 & 2020 Dec 18 - 05:48:42 &  108.7  \\
0882340601 & 2021 Apr 18 - 22:17:01 & 104.7 \\
0882340701 & 2021 Jul 17 - 18:11:06 & 83.5 \\
0882340801 & 2021 Oct 27 - 11:29:46 & 92.5 \\
\hline
\end{tabular}
\label{table_obs}
\end{table}
\section{Observations and Data Reduction}\label{sec:data}
\begin{figure}
\begin{center}
\includegraphics[width=\columnwidth]{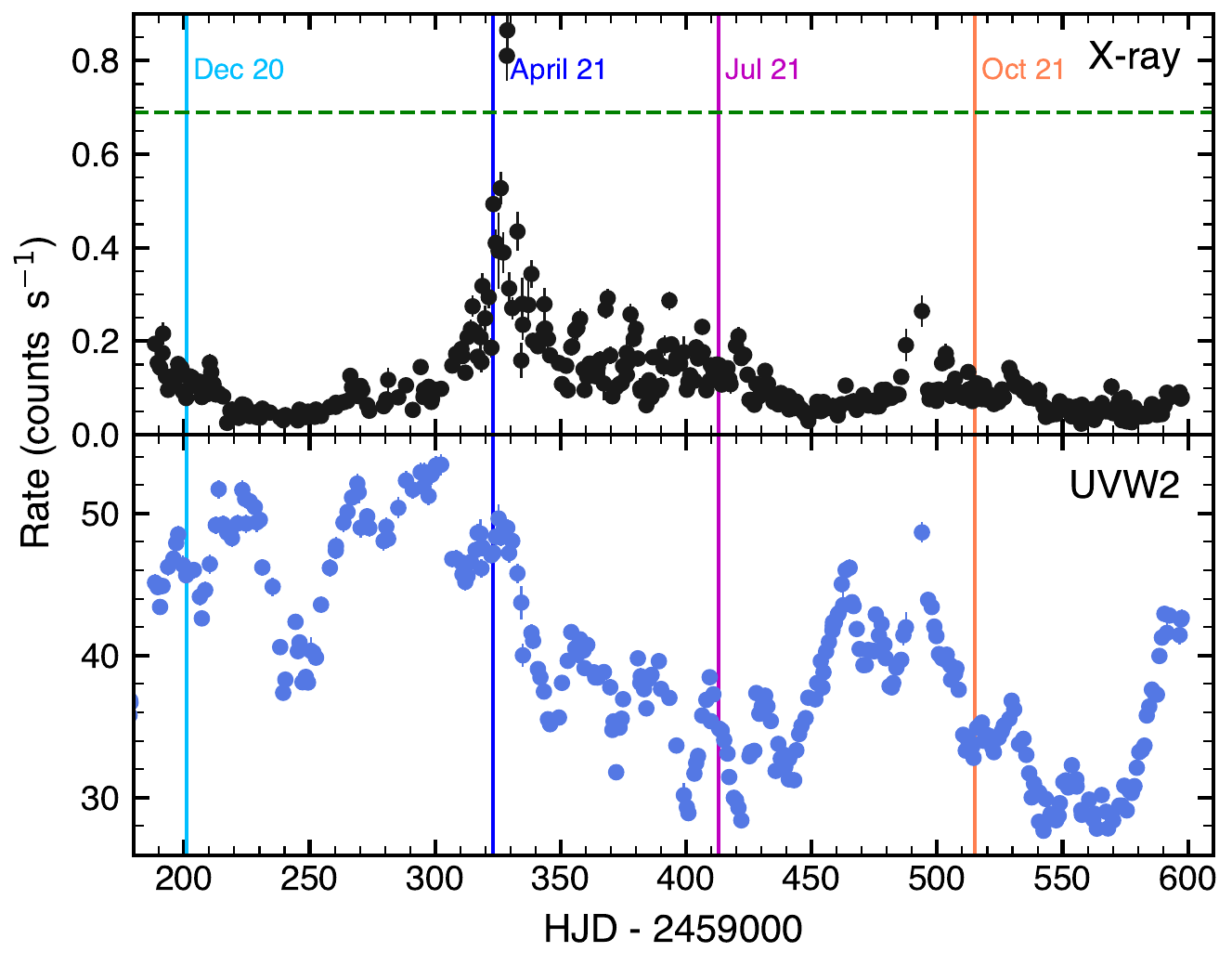}
\caption{ The AGN STORM 2 Swift XRT and UVW2 light curves from 2020~Dec to 2022~Feb (\citealt{2023arXiv230617663C}). Vertical lines indicate when the XMM-Newton and NuSTAR observations were conducted. The green horizontal line indicates the anticipated count rate based on the mean flux of an archival Swift campaign (\citealt{2019ApJ...870...54M}). During the AGN STORM 2 campaign, the source was largely obscured, with a mean Swift/XRT count rate below 0.2 counts~s$^{-1}$, except for a high flux state in 2021~Apr. Our second XMM-Newton and NuSTAR observation serendipitously took place during 2021~Apr.}

\label{fig:swift}
\end{center}
\end{figure}

There were four XMM-Newton with four simultaneous NuSTAR observations during the AGN STORM 2 campaign. One long 100~ks Chandra/HETG observation was also taken concurrent with the third XMM-Newton and NuSTAR epoch. Table \ref{table_obs} lists the start time and net exposure time of each XMM observation and Fig.~\ref{fig:swift}  shows the Swift XRT and UVW2 light curves from 2020 Dec to 2022 Feb, with vertical lines indicating when our four deep X-ray observations took place.

\begin{figure*}[!tbp]
\begin{center}
    \includegraphics[width=0.7\textwidth]{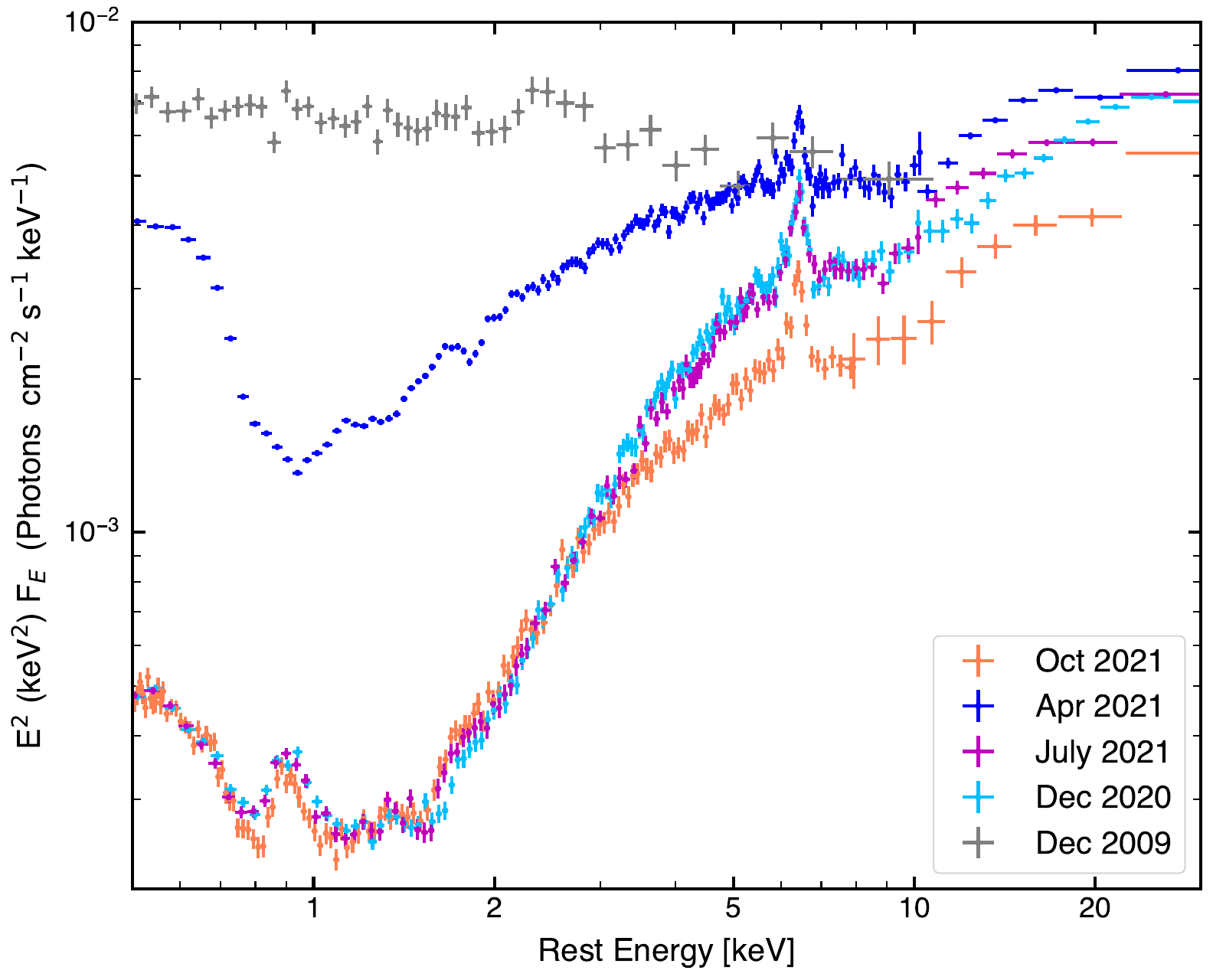}
    \caption{Flux density multiplied by $E^2$ for the four XMM-Newton+NuSTAR epochs from the prime AGN STORM 2 X-ray campaign. For comparison, we show the archival XMM-Newton observation from December 2009  (\citealt{2011ApJ...728...28W}) in gray. There is significant variability in the soft X-rays (due to the ionized obscurer) as well as changes in the hard X-ray emission indicative of intrinsic variability in the corona. The 2021~Apr observation, in blue, is the focus of this paper.}
    \label{fig:alldata}
\end{center}
\end{figure*}

Fig.~\ref{fig:alldata} shows the XMM-Newton pn and NuSTAR spectra along with the only archival XMM-Newton observation, a 15 ks observation taken in 2009 when the source was significantly brighter in the soft X-rays (\citealt{2011ApJ...728...28W}).  Our new XMM-Newton spectra show that there is significant long-term variability, particularly in the soft X-ray band (Fig.~\ref{fig:alldata}). This can largely be attributed to a change in obscuration (see Paper~1), however, the X-rays vary by a factor of $\sim 2$ even at high energies ($\sim 20$~keV), indicative of variability intrinsic to the X-ray corona. 
 
The second of our planned XMM-Newton and NuSTAR observations fortuitously took place during the only strong X-ray state of the entire campaign (2021~Apr in Fig.~\ref{fig:swift}). The NICER spectra revealed that this high flux state is not only due to a slight increase in ionizing luminosity, but also because the obscuring gas becomes more transparent due to an increase in ionization parameter and decrease in column density (\citealt{2023arXiv230212896P}). Our XMM-Newton observation from 2021~Apr has high enough signal to noise that we are able to see individual blueshifted absorption lines from the obscurer in the high resolution RGS spectra in addition to weak emission lines (see Fig.~\ref{fig:RGS}). The other three observations were so highly obscured that only emission lines from distant photoionized gas are observed and no absorption lines are detected (Fig.~\ref{fig:RGS}). 
The 2021~Apr observation and the properties of the obscuring outflow are the focus of this paper. The other three observations will be the subject of a future paper examining the nature of more distant, parsec-scale gas flows around the AGN.

\subsection{XMM-Newton}
We used the XMM-Newton Science Analysis System (SAS~v.~19.0.0) and the most recent calibration files to process the EPIC-pn data.  The observations were taken in Large Window Mode. We followed standard reduction procedures as outlined in the XMM Newton data reduction threads. The source extraction region was a 35 arcsec radius circular region centered on the source. The background regions are also 35 arcsec circular regions on the source detector avoiding contamination by the instrumental copper line at the edges of the detector. The count rate limit for flaring was set to 0.4 counts~s$^{-1}$ \textcolor{black}{based on the single event, high energy ($10-12$ keV)} pn background light curve. The observations were clean of soft proton background flares except at the edges of the observation's background light curve. The final exposure time was 104.7 ks. We produced the response matrices using {\sc rmfgen} and {\sc arfgen} in SAS. The EPIC-pn spectra were binned to have at least 25 counts for each background-subtracted spectral channel and without oversampling the intrinsic energy resolution by a factor larger than 3. We modeled the 1.5~keV~--~10~keV energy range in order to allow the RGS data to dominate the fits to the data at soft X-ray energies.
The reduced EPIC-pn spectra are shown in Fig.~\ref{fig:alldata}. Analysis of the Dec 2020 observation was presented in Paper~1, the 2021~Apr is the subject of this paper,  and the other observations will be the focus of a future paper.

\begin{figure*}[!tbp]
\begin{center}
\includegraphics[width=\textwidth]{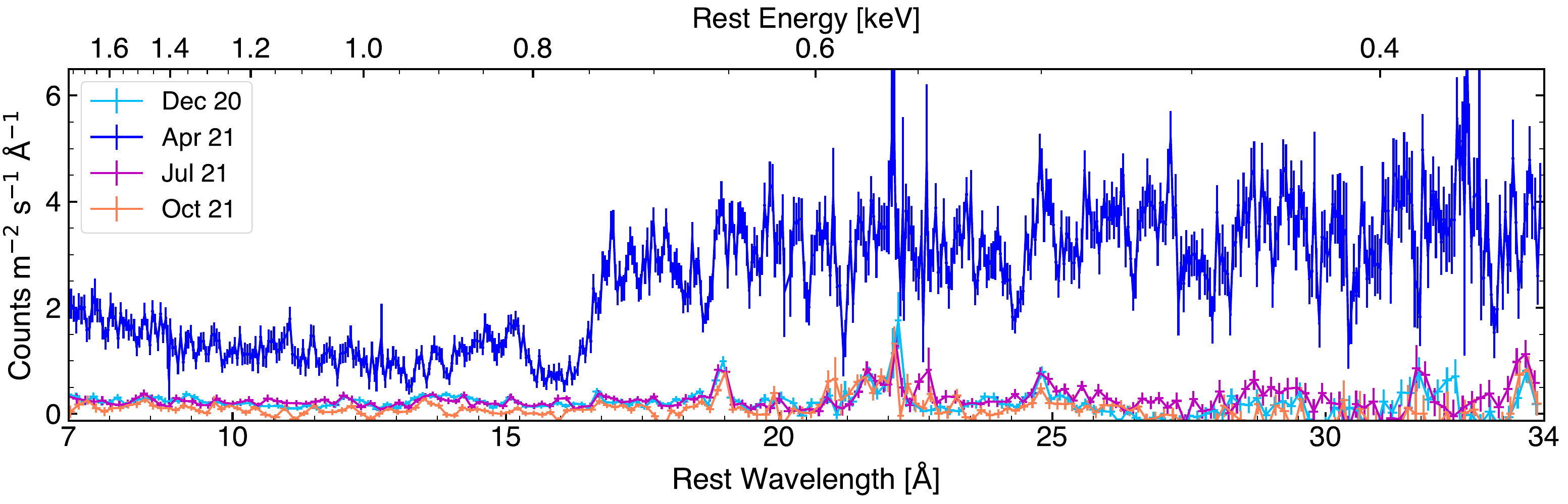}
\caption{The stacked RGS spectra from the four epochs. The 2021~Apr high-flux observation (blue) shows clear absorption troughs and is the subject of this paper. All the observations are binned used ‘optimal binning’ (\citealt{2016A&A...587A.151K}). The 2020~Dec, 2021~Jul, and 2021~Oct RGS spectra are shown in light blue, purple, and orange, respectively. These epochs are lower flux and show clear emission lines (Zaidouni et al., in prep.). These three observations are further binned by a factor of 3 for visualization purposes.}
\label{fig:RGS}
\end{center}
\end{figure*}

For the XMM-Newton RGS data, we use the pipeline tool {\sc rgsproc} to produce source and background spectral files, and instrument response files following standard procedures. We used 0.11 counts~s$^{-1}$ as the background flaring limit. The final exposure time is 129.5 and 129.2 ks for RGS 1 and 2, respectively. We bin the RGS 1 and 2 spectra using the ‘optimal binning’ from \cite{2016A&A...587A.151K}. We used the 7 {\AA} -- 35 {\AA} wavelength range of the RGS data in our fits (corresponding to 0.35 keV -- 1.77 keV). The reduced RGS data for the four epochs is shown in Fig.~\ref{fig:RGS}. The 2021~Apr RGS data show clear absorption troughs and have higher flux than the other three observations, which are dominated by emission lines.

\begin{figure}
\begin{center}
\includegraphics[width=\columnwidth]{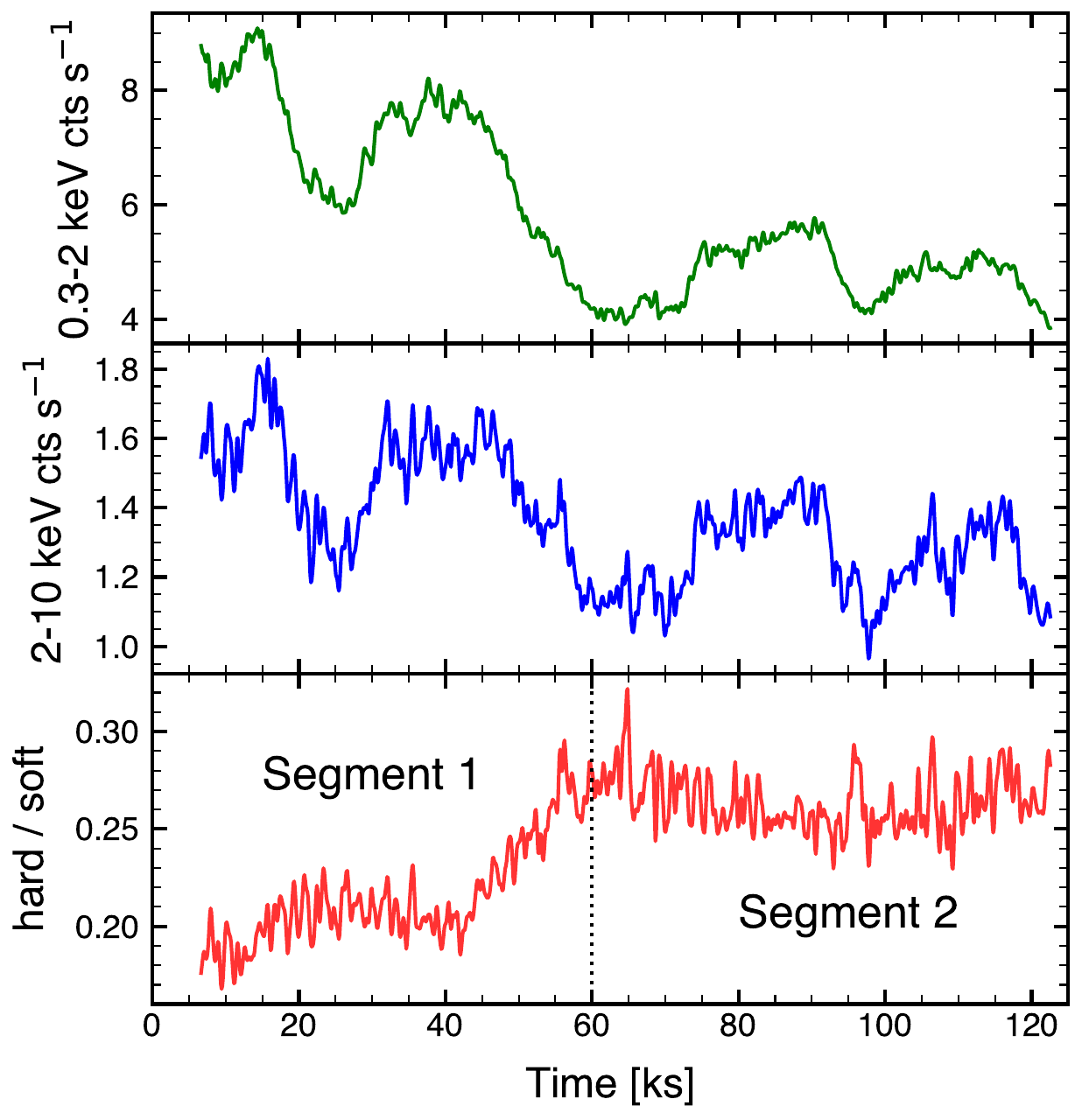}
\caption{ The XMM-Newton EPIC-pn soft band (0.3--2~keV; top panel) and hard band (2--10~keV; middle panel) light curves in 100~s bins along with the corresponding hardness ratio (hard/soft) for the 2021~Apr observation. The spectrum hardens significantly at $\sim 60$~ks, and the dotted black line delineates the division into two segments for the time-resolved analysis. 
}

\label{fig:lightCurves}
\end{center}
\end{figure}

Fig.~\ref{fig:lightCurves} shows the extracted EPIC-pn light curves in the soft (0.3--2~keV) and hard (2--10~keV) energy bands with the corresponding hardness ratio. Both the soft and hard X-ray bands clearly show rapid variability. Interestingly, the hardness ratio shows much less variability, but does show a 50\% increase (i.e., hardening) about halfway through the observation. As will be discussed in Section~\ref{subsec:timeResolved}, this change in spectral shape motivates us to analyze the RGS spectra in two separate segments.

\subsection{NuSTAR}

Concurrent NuSTAR observations were taken for each of the four epochs. We used the NuSTAR Data Analysis Software ({\sc nustardas v2.0}) to process the observations. We used the {\sc nupipeline} task and {\sc caldb} version 20100101v001 to produce and calibrate Level 2 event files. The source and background regions were circular regions with radii of 35 arcsec and 124 arcsec, respectively. We bin the data so that the signal-to-noise is greater than $3\sigma$ in each bin and oversample the instrumental resolution by a factor of three. We modeled the 3 -- 50 keV energy range. The NuSTAR spectra is shown in Fig.~\ref{fig:alldata}.

\begin{deluxetable*}{ccccc}
\label{tab:params}
\tablecaption{Model parameters}
\tablehead{
\colhead{Component} & \colhead{Parameter} & \colhead{Time-Averaged} & \colhead{Segment 1} & \colhead{Segment 2} }

\startdata
{\sc pow} &  $\Gamma$ & $2.0391 \pm 0.016$  & $2.085_{-0.014}^{+0.002}$ & $1.988_{-0.022}^{+0.023}$ 
\\ 
& Norm ($10^{49}$ ph s$^{-1}$ keV$^{-1}$)  & $1303_{-41}^{+54}$ & $1516 \pm 4$ & $1115_{-52}^{+54}$ 
 \\ 
\hline
{\sc bb} &  T (keV)  & $0.0876 \pm 0.0021$  & $0.0880 \pm 0.0003$ & $0.086_{-0.004}^{+0.003}$ 
\\ 
& Norm ($10^{16}$ m$^{2}$)  & $3800_{-400}^{+500}$ & $3900 \pm 100$ & $3500_{-600}^{+800}$ 
 \\ 
\hline
{\sc refl} &  $\xi$ (erg cm s$^{-1}$) &$0.05_{-0.01}^{+0.02}$ & $0.02 \pm 0.01$ &  $0.06_{-0.02}^{+0.04}$ 
\\ 
&  scal & $0.62 \pm 0.04 $ & $0.62_{-0.02}^{+0.03}$ & $0.60 \pm 0.06$ 
 \\ 
\hline
{\sc pion 1} & $N_\mathrm{H}$ ($10^{21}$ cm$^{-2}$)  & $22.346 \pm 0.001$ & $20.42_{-0.19}^{+0.18}$ & $25.11_{-1.69}^{+1.74}$ 
 \\
&  $\log \xi$ (erg cm s$^{-1}$)& $2.31_{-0.02}^{+0.03}$ & $2.33 \pm 0.01$ & $2.40 \pm 0.04$  
\\
& Cov. Frac.~$\Omega$ & $0.860_{-0.072}^{+0.005}$ & $0.841 \pm 0.004$ & $0.898 \pm 0.012$ 
\\ 
& $\sigma_v$ (km s$^{-1}$)  & $93 \pm 11$ & $116_{-10}^{+11}$ & $80_{-13}^{+15}$ 
\\ 
& $v_{\text {out }}$ (km s$^{-1}$)  & $-5900 \pm 40$ & $-5800 \pm 50$ & $-5900 \pm 50$ 
\\ 
\hline
{\sc pion 2} & $N_\mathrm{H}$ ($10^{21}$ cm$^{-2}$)  & $42.485 \pm 0.005$ & $42.48_{-0.59}^{+4.93}$ & $28.57_{-6.18}^{+7.05}$ 
\\  
&  $\log \xi$ (erg cm s$^{-1}$)& $3.46 \pm 0.03$ & $3.25 \pm 0.01$ & $3.34 \pm 0.06$ 
\\ 
& Cov. Frac.~$\Omega$ & 1\tablenotemark{a} & $> 0.96$ & 1\tablenotemark{a} \\ 
& $\sigma_v$ (km s$^{-1}$)  & $67_{-13}^{+14}$ & $69_{-5}^{+6}$ & $128_{-30}^{+35}$ 
 \\ 
& $v_{\text {out }}$ (km s$^{-1}$)  & $-5500_{-40}^{+60}$ & $-5400 \pm 50$ & $-5600 \pm 80$ 
\\  
\hline
{\sc pion 3} & $N_\mathrm{H}$ ($10^{21}$ cm$^{-2}$) & $1.18_{-0.37}^{+0.39}$ & $0.68_{-0.1}^{+0.88}$ & 
\\  
&  $\log \xi$ (erg cm s$^{-1}$) & $2.90_{-0.06}^{+0.05}$ & $2.68_{-0.07}^{+0.06}$ &
\\ 
& Cov. Frac.~$\Omega$ & 1\tablenotemark{a} & $> 0.75$ & --- \\
& $\sigma_v$ (km s$^{-1}$) & $73_{-22}^{+31}$ & $157_{-36}^{+49}$ &  
\\  
& $v_{\text {out }}$ (km s$^{-1}$)  & $-3900 \pm 100$ & $-4100 \pm 100$ &   \\
\hline
 & $\chi^2$/d.o.f. & 1895/1532 & 1723/1520 & 1756/1508\\
\enddata
\tablenotetext{a}{Fixed parameter}
\end{deluxetable*}

\section{Results}\label{sec:results}
We analyze the 2021~Apr observation using the X-ray spectral fitting package {\sc spex} (\citealt{1996uxsa.conf..411K}), which allows us to investigate the photoionization structure of the outflowing gas producing the absorption lines.  Throughout our analysis, we used \cite{1979ApJ...228..939C} statistics to assess the goodness of fit and the errors reported are within $1\sigma$ errors. Distance calculations assume $H_0 =  70.0$ km~s$^{-1}$~Mpc$^{-1}$, $\Omega_m = 0.300$, $\Omega_\Lambda = 0.700$, and $\Omega_r = 0.000$ leading to a luminosity distance of 137.96 Mpc (\citealt{1988AJ.....95.1602S}).

\subsection{Model Components}\label{subsec:model-comps}

\begin{figure*}[!tbp]
\begin{center}
\includegraphics[scale=0.5]{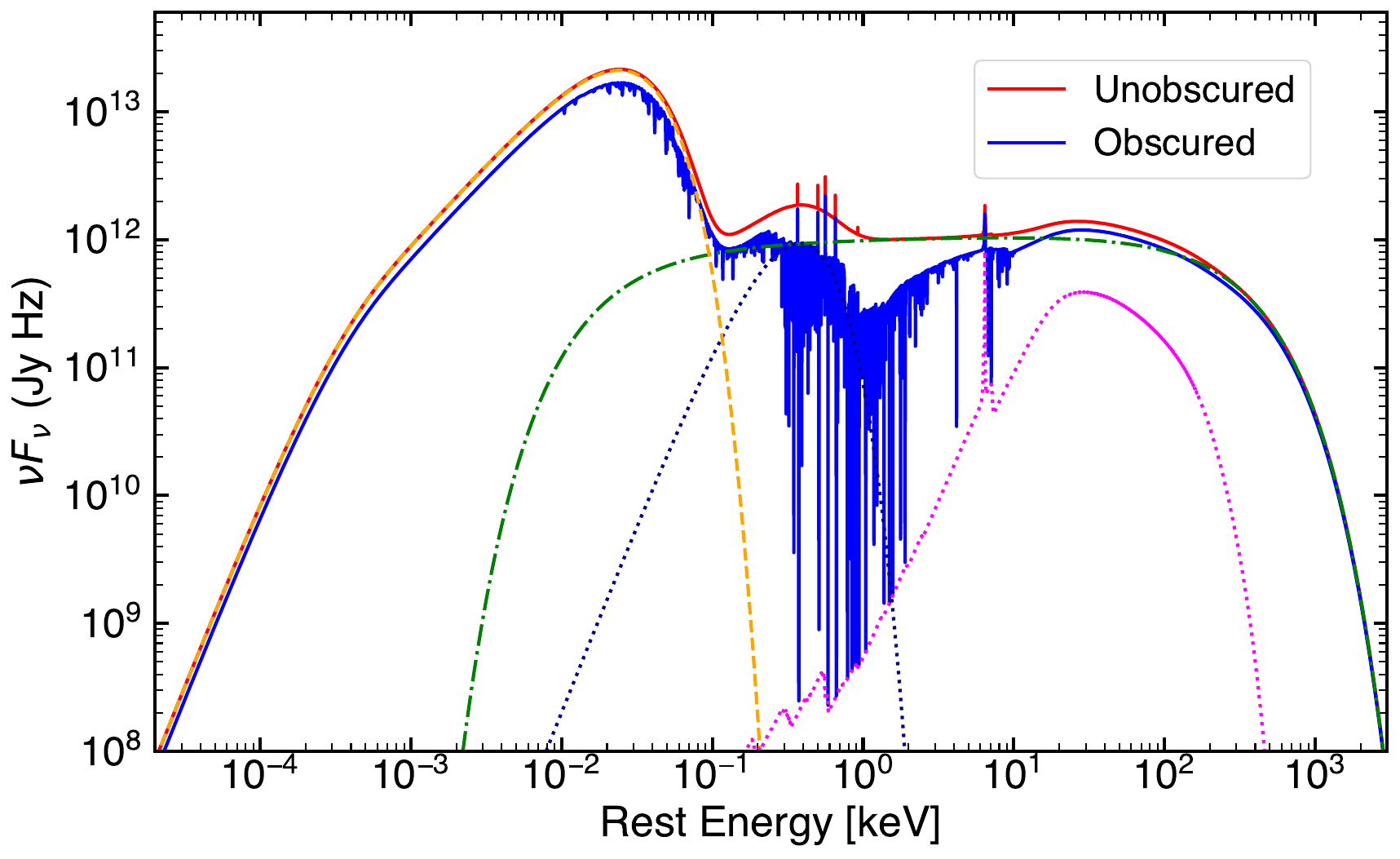}
\caption{The best-fitting spectral energy distribution (SED) model. The total obscured (observed) SED is shown in blue, and the inferred unobscured (intrinsic) continuum in red. The model components are shown using non-solid lines. The X-ray continuum is modeled as a cut-off powerlaw (green dot-dash). We model the soft excess as a phenomenological black body spectrum (blue dotted). The accretion disk is modeled as a multi-temperature disk blackbody (orange dashed). The X-ray reflection (iron K$\alpha$ and Compton hump emission) is shown as a pink dotted line.}
\label{fig:SED}
\end{center}
\end{figure*}

\begin{figure*}[!tbp]
\begin{center}
    \includegraphics[width=\textwidth]{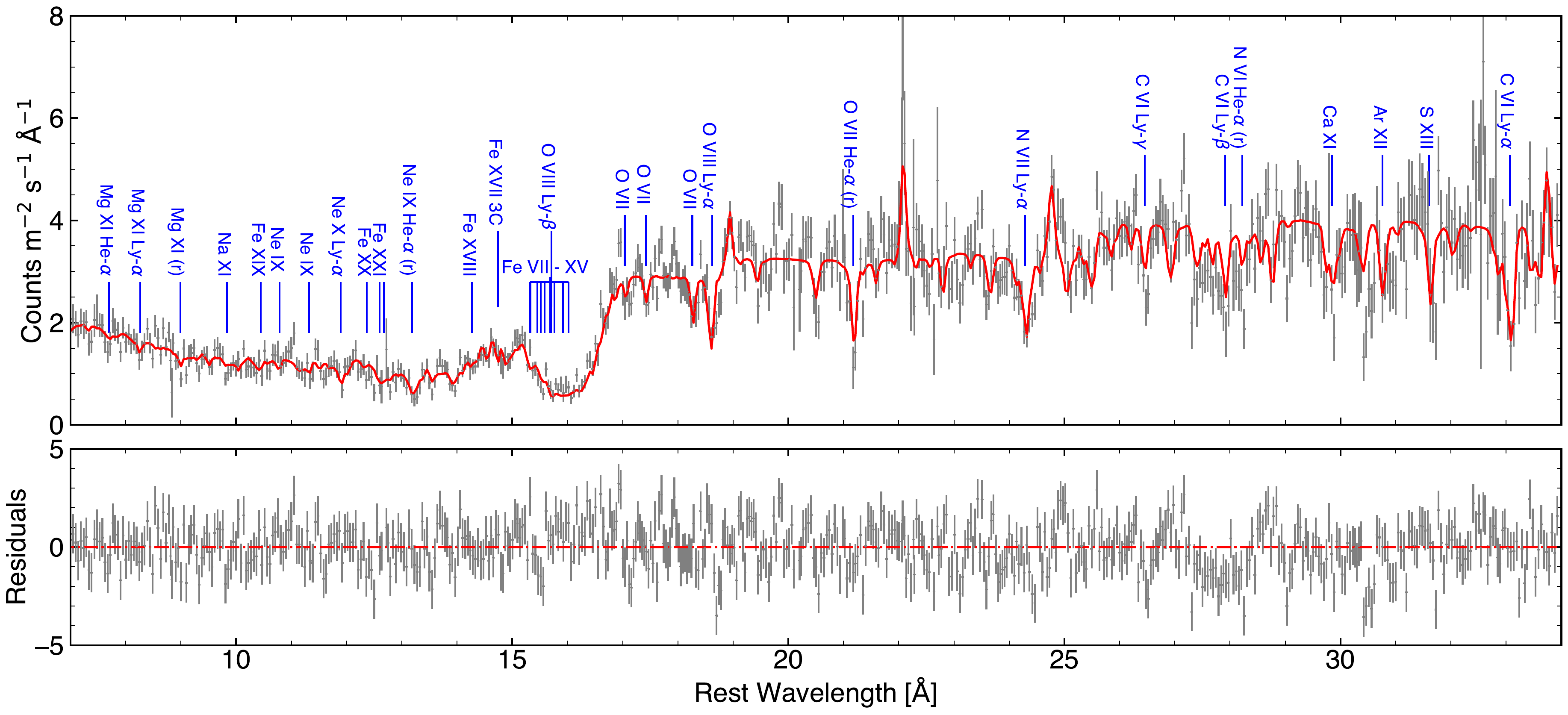}
    \caption{The time-averaged RGS spectrum with the best fitting model overlaid in red. A subset of the absorption lines are labeled at their blue-shifted wavelengths. The residuals are calculated as “(model-data)/error”.}
\label{fig:RGS_overall}
\end{center}
\end{figure*}

\begin{figure*}[!tbp]
\begin{center}
\includegraphics[width=\textwidth]{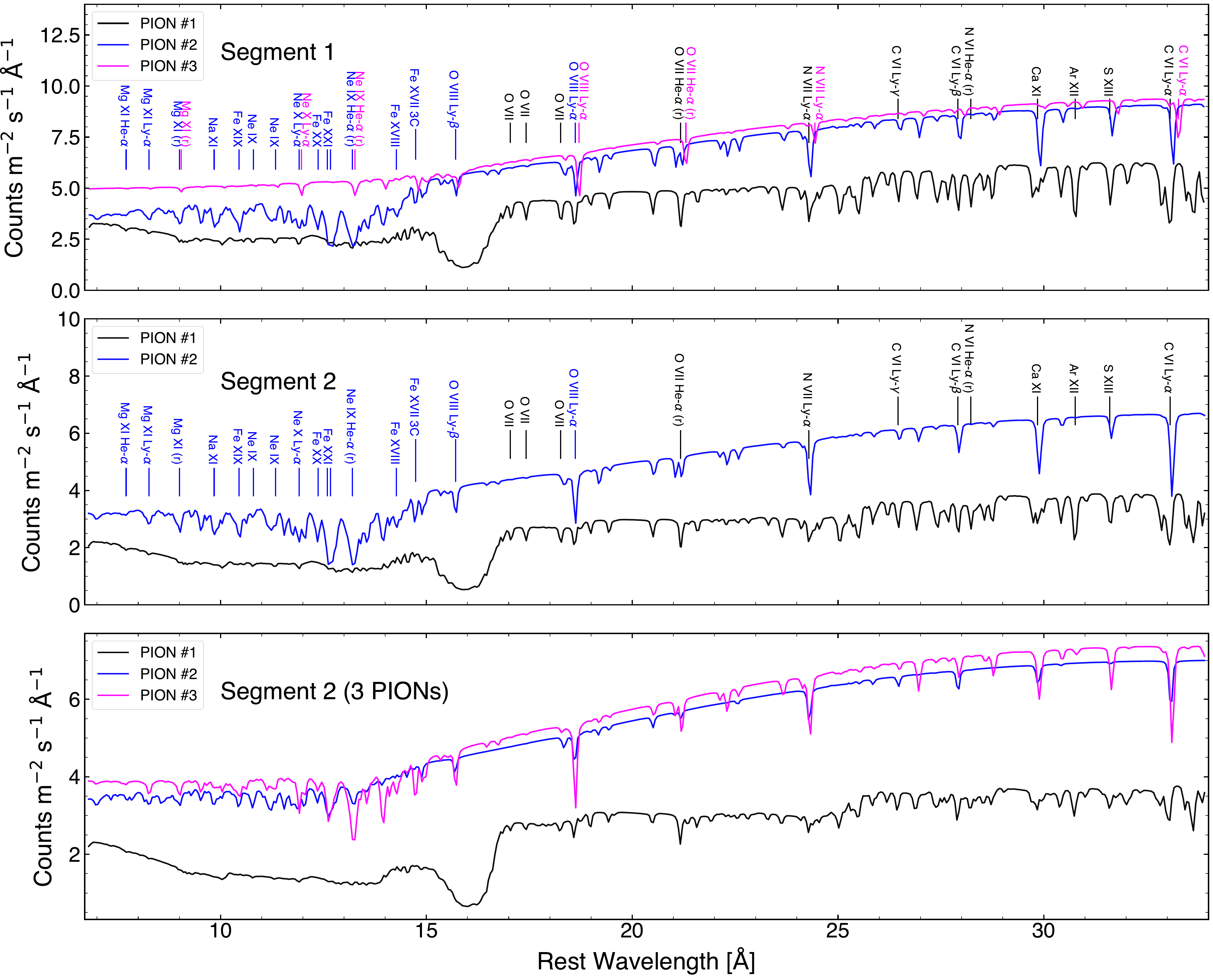}
\caption{The contribution of each of the {\sc pion} photoionization model components to the overall fit for Segment 1 (t~$<60$~ks, top panel), and Segment 2 (t~$>60$~ks, middle panel). The bottom panel shows the scenario where a statistically significant, but redundant {\sc pion} component is added to the fit of Segment 2. The models are in the form of “${\tt (pow\times etau \times etau + bb + dbb + refl) \times {\sc pion}}$” which we achieve by setting the covering fraction of the other {\sc pion} components to zero. For both Segments 1 and 2, {\sc pion} 1 (black) is the lowest ionization component, fitting largely the long wavelength lines and the iron UTA, while {\sc pion} 2 (blue) is a higher ionization component, modeling largely the shorter wavelength lines. {\sc pion} 3 (pink) accounts for the asymmetry in the line shapes of Segment 1 (see text for details).}

\label{fig:contrib}
\end{center}
\end{figure*}

\begin{figure}[!tbp]
\begin{center}
\includegraphics[width=\columnwidth]{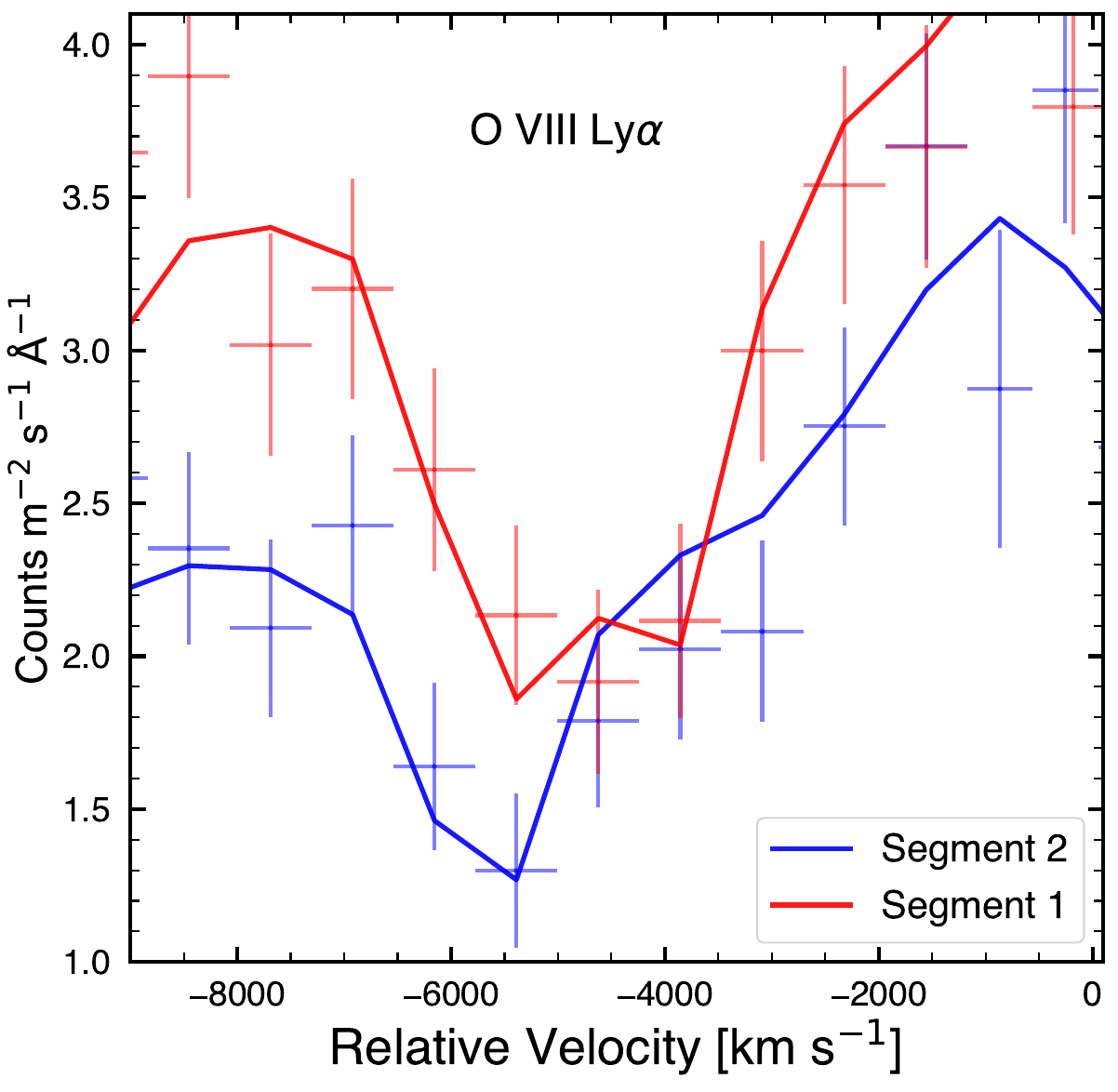}
\caption{ The O VIII Ly$\alpha$ from Segment 1 (t~$<60$~ks, red points) and Segment 2 (t~$>60$~ks, blue points) with the best-fitting models. This is an example of the enhanced red wing present in Segment 1 and not Segment 2.}
\label{fig:zoom}
\end{center}
\end{figure}

\begin{figure}
\begin{center}
\includegraphics[width=\columnwidth]{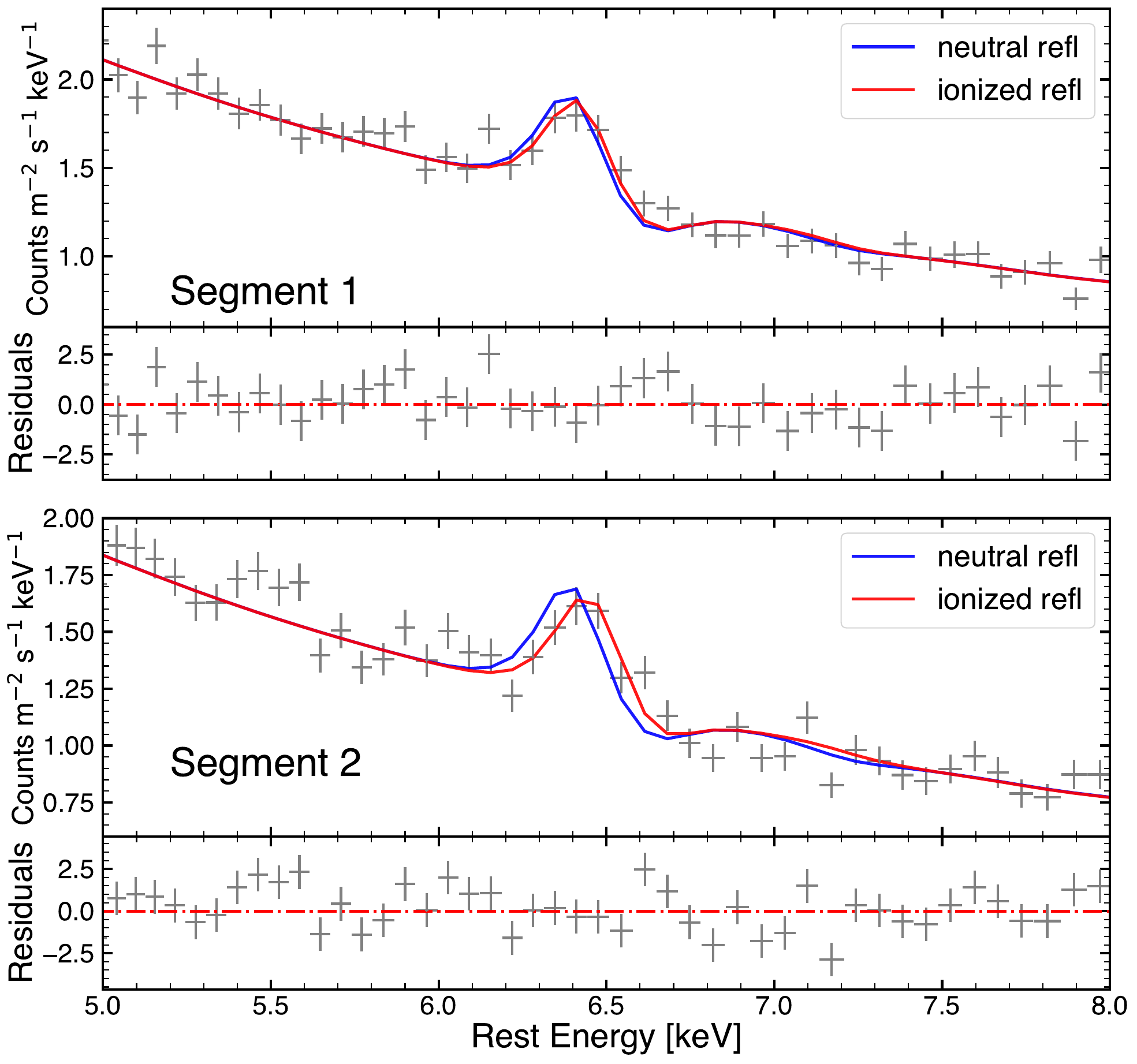}
\caption{The region around the Fe K alpha emission line from the EPIC-pn instrument data. We overlay the best fit model in red. There are no strong absorption features in this energy band, but we find that the data prefer a mildly ionized reflection model over a neutral reflection model (blue), particularly in Segment~2 when the spectrum is harder (see Table \ref{tab:params} for the parameter values).}
\label{fig:Fe}
\end{center}
\end{figure}

To construct a model for the ionizing SED, we fit the XMM-Newton + NuSTAR observation taken in 2021~Apr. The broadband SED is needed to accurately determine the ionization balance for the photoionized absorber model. The EPIC-pn and NuSTAR data constrain the lower-resolution broadband spectra ($1.5-50$ keV), while the RGS data cover the $10-36$ {\AA} ($0.35-1.2$ keV) range. Similar to previous analyses (e.g. \citealt{2014Sci...345...64K}), we omit the 0.3--1.5 keV pn data from our analysis since we do not want the higher count-rate CCD spectra to dominate the model fits over the RGS spectra.

The best-fitting obscured and inferred unobscured (continuum) SED models are shown in Fig.~\ref{fig:SED}. In this section we describe the different components of the broadband continuum model.

\begin{itemize}

\item Thermal emission from the accretion disk peaks in the UV, and we model this as a multi-temperature disk component (DBB in {\sc spex}, orange dashed line in Fig~\ref{fig:SED}). The temperature and normalization are fixed to the values from fits to the STIS+COS NIR to UV spectrum presented in Paper~1.

\item The X-ray continuum is produced by inverse Compton scattering of disk photons in the mildly relativistic corona. We model the X-ray continuum as a cut-off powerlaw ({\tt pow} and {\tt etau} in {\sc spex}, green dot-dash in Fig.~\ref{fig:SED}). The upper energy is fixed at 300~keV, typical of AGN X-ray coronae (\citealt{2015MNRAS.451.4375F}). We set a lower energy cut-off of 20 eV (equivalent to the maximum temperature of the disk component). Freeing the energy limits does not significantly change the fit.

\item Paper~1 showed that the inferred outflow properties were insensitive to the choice of physical model for the soft excess (i.e. a `warm corona', e.g. \citealt{2018A&A...611A..59P}, or blurred reflection, e.g. \citealt{2013ApJ...768..146G}), so we simply model the soft excess as a blackbody ({\tt bb} in {\sc spex}, blue dotted line in Fig~\ref{fig:SED}).

\item We model distant X-ray reflection (pink dotted in Fig.~\ref{fig:SED}) using the {\tt refl} model (non-relativistic reflection) which accounts for a narrow iron K-line and the Compton hump (\citealt{1995MNRAS.273..837M,1999MNRAS.305..231Z}). The {\tt refl} model in {\sc spex} is a relatively basic reflection model  which suffices for the current analyis. We will consider more detailed reflection models later, as in Paper 1 and \cite{2021ApJ...911L..12M}.

\item We use the {\tt hot} model in {\sc spex} to account for Galactic absorption with a fixed temperature of \mbox{$8\times10^{-6}$}~keV corresponding to the case of cold ISM plasma. We fix the Galactic column density to $N$(\ion{H}{1})$ =~1.15~\times~10^{20}~$cm$^{-2}$  \citep{Murphy96}.

\begin{figure}[]
\begin{center}
\hspace*{-1cm}
\includegraphics[scale=0.42]{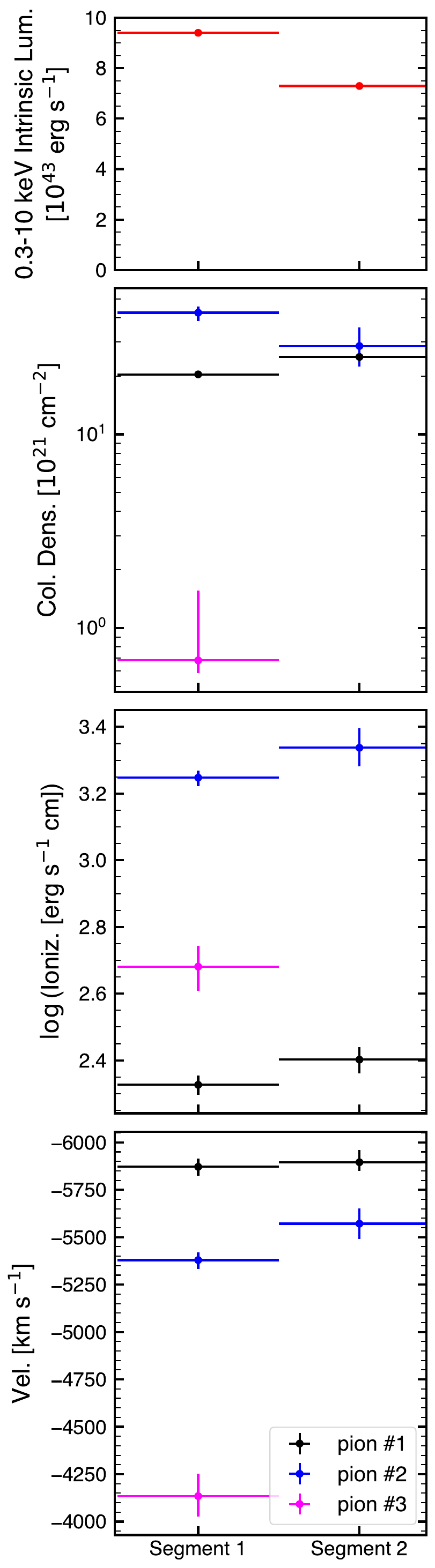}
\caption{The evolution of key obscurer parameters between the first and second half of the observation. The top panel shows the intrinsic luminosity (absorption corrected) between $0.3$ and $10$ keV. The rest of the panels show the best-fitting values of the column density, ionization, and outflow velocity for each segment (values from Table \ref{tab:params}).}
\label{fig:param_evol}
\end{center}
\end{figure}

\item The ionized obscurer is modeled using the photoionization model {\sc pion}. {\sc pion} uses the ionizing spectrum from the continuum components to calculate the transmission and emission of a slab of photoionized plasma. It takes into account the time variability of the continuum in the calculations of the ionization balance (\citealt{2016A&A...596A..65M,2015Natur.526..542M}). The abundances are set to the proto-Solar abundances of \citet{2009LanB...4B..712L}. We find that the RGS data statistically require 3 {\sc pion} components to fully describe the shape of the absorption lines. Given the degeneracy between the covering fraction and column density, constraining both parameters simultaneously is challenging, so we often fix the covering fraction at 1 to simplify the analysis. More details about the photoionization modeling are presented in the time-resolved analysis section (Section~\ref{subsec:timeResolved}).

\item Analysis of the XMM+NuSTAR X-ray spectrum from December 2020 (obs ID: 0872390901, Table \ref{table_obs}) in Paper~1
reveal emission from distant circumnuclear gas outflowing with a velocity of $v=-400\pm70$\ km~s$^{-1}$. Although the 2021~Apr RGS observation is dominated by strong, narrow absorption lines, it also shows signatures of these narrow emission lines so we add Gaussian profiles ({\tt gaus} in {\sc spex}, \citealt{1809tmcc.book.....G}) for the five detected emission lines: Ne~IX~He-$\alpha$ ($13.45$~\AA), O~VIII~Ly$\alpha$ ($18.97$~\AA), O~VII~(f) ($22.101$~\AA), N~VII~Ly$\alpha$ ($24.78$~\AA), and C~VI~Ly$\alpha$ ($33.74$~\AA). The line wavelengths are fixed to their rest energies, the line full width at half maxima (fwhm) are coupled between the emission lines, and the normalizations were left as free parameters. A single {\tt reds} component was added to account for the velocity shift in all 5 lines.
\end{itemize} 
Hence, the full model in {\sc spex} syntax is:
\begin{equation}\label{equ:final-model}
  \begin{array}{l}
    {\tt 
    \{(pow\times 2etau + bb + dbb)\times3pion + refl} \\ 
    {\tt + 5gaus\times reds \} \times hot\times reds }
  \end{array}
\end{equation}

Finally, we searched for the UFO previously discovered in Mrk~817 by \citet{2023arXiv231206487Z}. They observe a structured ultra-fast outflow in Mrk 817 using XMM-Newton and NuSTAR observations obtained during a low flux state in April 2022. Motivated by their result, we searched for a blueshifted Fe~XXVI UFO line (rest frame energy of 6.97 keV) using a Gaussian line scan of the iron line region. We allow the blueshift to vary between $0.039c$ and $0.099$c. This range is chosen given the uncertainties bounding the velocities seen in \citet{2023arXiv231206487Z} in addition to at least $\pm 10\%$ of the uncertainty. We check using a full width at a half maximum of 0 km s$^{-1}$ and 1000 km s$^{-1}$. We find a shallow feature blueshifted at $v / c = -0.039_{-0.0001}^{+0.002} $. The fit improved by $\Delta C = 5$ for two d.o.f., corresponding to an improvement at the $1.74$ sigma level and thus, it is not significantly detected. We will explore this in our future analysis of
the low state observations.

The blue line in Fig.~\ref{fig:SED} shows the final model, after the continuum is obscured. Fig.~\ref{fig:RGS_overall} shows the RGS spectrum and the best fitting model.  The unobscured SED has a $1-1000$ Ryd ($0.0136-13.6$ keV) ionizing luminosity of $8.1 \times 10^{44}$ erg~s$^{-1}$ compared to $6.5 \times 10^{44}$ erg~s$^{-1}$ for the obscured SED.

\subsection{Time-resolved Analysis}\label{subsec:timeResolved}

Motivated by the change of the hardness ratio and count rate in the pn lightcurve shown in Fig.~\ref{fig:lightCurves}, we split the light curve into two segments (Segment 1: $t <$ 60~ks and Segment 2: $t >$ 60~ks) as indicated by the dotted black line. We reduced each segment following the same procedures outlined in Section~\ref{sec:data} for the RGS, XMM pn and NuSTAR data.  The time-resolved RGS spectra (shown in Fig.~\ref{fig:RGS_segments} in the appendix) show changes in the absorption lines. For example, the absorption troughs in Segment 2 are generally narrower than in Segment 1 (see the close-up of the O VIII Ly$\alpha$ line in Fig.~\ref{fig:zoom}). This indicates that in addition to changes in the absorption and continuum observed in the broadband data between observations (Fig.~\ref{fig:alldata}), the outflow properties also change on timescales of $<$ 1 day. This further motivates a time-resolved modeling analysis.

In what follows, “component” refers to each of the photoionization models ({\sc pion} in {\sc spex}) used in fitting the RGS absorption lines, while “segment” refers to the time-resolved spectra obtained by splitting the XMM-Newton (pn+RGS) + NuSTAR observations into two segments.

Using the model in Equation \ref{equ:final-model}, we fit the RGS, pn and NuSTAR spectra of Segment 1 and Segment 2 separately, allowing the continuum to vary between segments. 

The final model fits for both segments are shown in Fig.~\ref{fig:RGS_segments}. Fig.~\ref{fig:Fe} shows a zoom-in of the pn spectra over the 5--8~keV range. No significant absorption lines (e.g. from Fe XXV or Fe XXVI) are predicted by the model or present in the data. The only major difference between Segments 1 and 2 in this band is in the shift in the iron emission line. In Segment~1, the model is largely consistent with neutral reflection, but in Segment~2, neutral reflection is not sufficient, and we find an ionization parameter of $0.05_{-0.01}^{+0.02}$ erg cm s$^{-1}$. The emission lines (both neutral and ionized) will be investigated in future works. 

The observed flux (0.3 keV to 50 keV) decreases by 32\% from Segment 1 to 2 and this decrease in flux and spectral hardening (Fig.~\ref{fig:lightCurves}) is due to an intrinsic luminosity decrease, rather than a significant increase in the overall obscuration (Fig.~\ref{fig:param_evol}).

Table \ref{tab:params} provides the final model parameters for the time-averaged spectrum and for each segment. Statistically, the final fit for Segment 1 requires three {\sc pion} photoionization components ($\Delta C = 228$ for 4 d.o.fs between a model with 1 {\sc pion} and 2 {\sc pions}, and $\Delta C = 35$ for 4 d.o.fs between a model with 2 {\sc pions} and 3 {\sc pions}, corresponding to an F-test significance of $> 8\sigma$ and $5\sigma$, respectively). On the other hand, the final fit for Segment 2 requires only 2 {\sc pion} components ($\Delta C = 136$ for 4 d.o.fs between a model with 1 {\sc pion} and 2 {\sc pions} corresponding to a significance $> 8\sigma$, and $\Delta C = 14$ for 4 d.o.fs between a model with 2 {\sc pions} and 3 {\sc pions}, corresponding to only a $2.6\sigma$ significance). There is a fitting scenario where adding a third {\sc pion} to Segment 2 improves the fit more significantly (by $5\sigma$) but, in that case, the {\sc pion} component has a very high column density ($N_{\mathrm{H}} > 3 \times 10^{23}~\text{cm}^{-2}$) and ionization parameter ($\log\xi>4$) in favor for a much lower reflection scale (scal $<0.4$). This would make it significantly Compton thick, likely requiring a strong re-emission component, which is not observed. A similar component shows up in Segment 1 upon adding a 4th {\sc pion}, so this scenario is not unique to Segment 2. Given that the neutral reflection component is not expected to change dramatically on such short timescales, and that this {\sc pion} component does not contribute to modeling any additional spectral lines (as detailed below and in Fig.~\ref{fig:contrib}), we conclude that the significance of the third component is due to a degeneracy between the parameters and not physical.

We visualize the contribution of each {\sc pion} component to the final fit in Fig.~\ref{fig:contrib}. In both segments, the first {\sc pion} component primarily fits the low ionization transitions including the unresolved Fe transition array (UTA, \citealt{2001ApJ...563..497B}), while the second {\sc pion} component primarily describes the higher ionization transitions. The third {\sc pion} component, present only in segment 1, models the red wing of the absorption troughs seen in prominent features such as C~VI~Ly$\alpha$ at a rest wavelength of $33.74$~\AA~and O~VIII~Ly$\alpha$ at a rest wavelength of $18.97$~\AA. A close-up of O~VIII~Ly$\alpha$ is shown in Fig.~\ref{fig:zoom}.

Our analysis shows that in Segment 1, the covering fractions for {\sc pion} 2 and 3 are both consistent with unity, as detailed in Table \ref{tab:params}. For Segment 2, to facilitate a direct comparison with Segment 1, we set the covering fraction of {\sc pion} 2 to 1 (otherwise a lower covering fraction can be found in favor of a higher column density). 

Most of the model parameters are consistent to within $2\sigma$ between Segments 1 and 2, specifically the ionization parameters, velocity broadening parameters, and the outflow velocities for {\sc pion} 1 and 2, along with the column densities of {\sc pion} 2. Conversely, the column densities and covering fractions of {\sc pion} 1 in Segment 2 exceed those in Segment 1, with approximately $3\sigma$ significance.

The best-fitting parameters, for both segments, are visualized in Fig.~\ref{fig:param_evol}. The obscurer is a multiphase outflow in terms of both ionization parameters and outflow velocities. It is also variable across the duration of the observation as the absorption increases and the intrinsic luminosity decreases. Additional variability is manifested by the slower photoionization component, responsible for the extended red wing in the first segment that is much less evident in the second segment.

\subsection{Comparison with the UV band}\label{sec:UV}

\begin{figure}[]
\begin{center}
\includegraphics[width=\columnwidth]{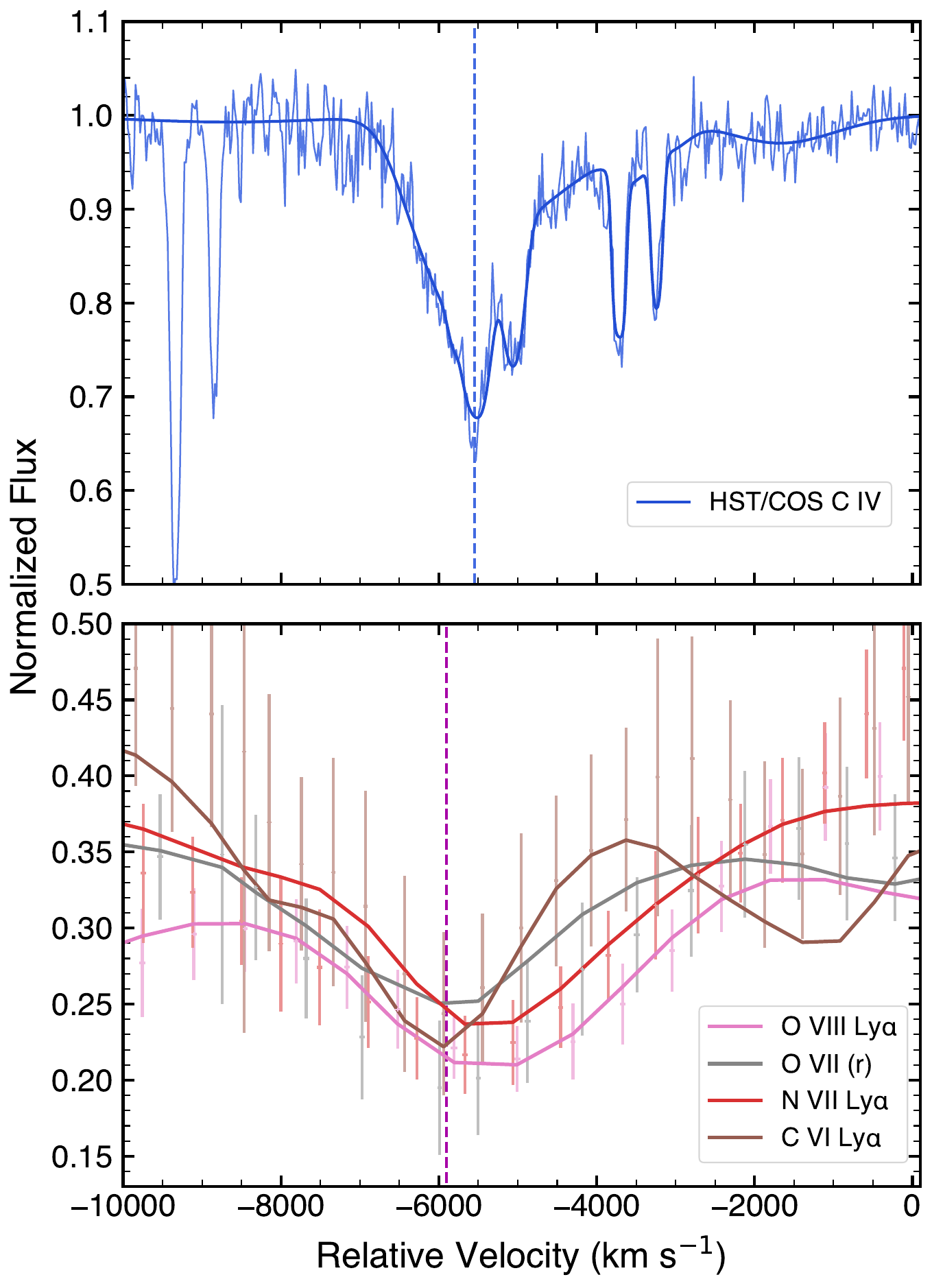}
\caption{The top panel shows the C IV line profile from an HST COS spectrum taken a day before the X-ray observation (see text for details). The bottom panel shows strong X-ray absorption line profiles from the overall 2021~Apr RGS spectrum, notably, the transitions: O VII Ly$\alpha$, O VII (r), N VII Ly$\alpha$, and C VI Ly$\alpha$ along with the model fit to the overall observation. The velocities are with respect to the rest frame of the source. The y-axis is the normalized flux with respect to the unabsorbed continuum. Both the X-ray and UV data show evidence for an obscurer outflowing at between $\sim -5000$ and $\sim -6000$ km~s$^{-1}$. The vertical dashed lines show the approximate centers of the absorption troughs.}
\label{fig:UV}
\end{center}
\end{figure}

In earlier AGN STORM 2 papers, we noticed broad UV absorption troughs in absorption lines such as C~IV and Si IV. These absorption troughs are blueshifted by $\sim 5000$ km~s$^{-1}$, and the equivalent widths (EW) of these troughs are correlated with the column density of the obscurer (\citealt{2023arXiv230212896P}, Paper~1). 
This co-evolution of the X-ray and UV absorption features suggests we are seeing an evolution of different ionization states of the same gas. The high-resolution X-ray spectra allow us to identify individual absorption lines so we can compare the dynamics and energetics of the UV and X-ray outflow in more detail.

Using the time-averaged RGS X-ray spectrum, we zoom in on several absorption lines to show their rest frame velocity profiles in Fig.~\ref{fig:UV}.
The X-ray absorption lines show a blueshift of $\sim 6000$ km~s$^{-1}$ matching the properties and velocities seen in the pion models (see Table \ref{tab:params}).

The top panel of Fig.~\ref{fig:UV} shows the C IV (1548.195 \AA) broad UV absorption troughs in the HST COS data from HJD=59322.4023 \textcolor{black}{(this data may be obtained from the MAST archive at\dataset[doi:10.17909/10sp-zt74]{https://dx.doi.org/10.17909/10sp-zt74})}.
The C IV absorption trough was chosen since it shows the kinematic structure of the outflow and is not blended with other transitions (Kriss et al. in prep). 

To fit the C IV absorption line, we start with the model of the emission spectrum as described in Paper~1. The absorption is complex, and we characterize it with several Gaussians. Each Gaussian has a freely varying strength, velocity, and width. The two components of the CIV doublet have their optical depths fixed at a 2:1 ratio (blue:red), velocities tied by the ratio of their rest wavelengths, and identical widths. The C IV fits include a narrow doublet at $\sim -3200~~\rm km~s^{-1}$, and a blend of four overlapping components that comprise the main portion of the absorption trough at $\sim-5500~~\rm km~s^{-1}$. This main portion of the absorption trough also has its overall profile shaped by another Gaussian that specifies a smoothly varying covering fraction. Thus, at lower optical depths when the individual lines are not saturated, one can begin to resolve the individual velocity components. The two strong lines at $\sim-9000$ km~s$^{-1}$ that are not part of the fit are the foreground interstellar C IV absorption lines. Details on the fits for the overall HST COS spectra of Mrk~817 will be presented by Kriss et al. in prep.

The most important point in our comparison of the X-ray RGS and the UV C IV absorption troughs is that both have a blueshifted component at $\sim -5500$ km~s$^{-1}$, indicating that they are likely from the same material. We discuss the interpretation of this finding in more detail in Section~\ref{sec:discussion}.

\section{Discussion}\label{sec:discussion}

In this paper we present the results on the photoionization and kinematic properties of the obscuring wind in Mrk~817. It has been shown to be highly variable (\citealt{2023arXiv230212896P}), and appears to have a significant affect on the measured time lags (both the broad line region lags; \citealt{2023arXiv230211587H}, and the continuum lags; \citealt{2023arXiv230617663C}, Lewin et al. in prep). We have 4 XMM observations at different points in the campaign (Fig.~\ref{fig:swift}). Fig.~\ref{fig:alldata} shows that the amount of obscuration correlates with the intrinsic luminosity. At one particularly bright observation during 2021-04-18, we were able to see X-ray absorption lines at 5000-6000 km s$^{-1}$.

Similar ionized obscurers have been seen before in NGC 5548 (\citealt{2014Sci...345...64K}), NGC 3227 (\citealt{2003MNRAS.342L..41L}), NGC 985 (\citealt{2016A&A...586A..72E}), NGC 3783 (\citealt{2017A&A...607A..28M}), MR 2251-17 (\citealt{1984ApJ...281...90H,2022ApJ...940...41M}), NGC 6814 (\citealt{1994ApJ...421...69L}), NGC 3516 (\citealt{2008A&A...483..161T}), and Mrk 335 (\citealt{2013ApJ...766..104L}). Typically, they have column densities of $N_{\mathrm{H}} \sim$ (0.2~--~200)$\times 10^{22} ~\mathrm{cm}^{-2}$, ionization parameters of $\log\xi \sim -2$ --~4  erg cm s$^{-1}$, covering fractions from 20\% to 100\%, and outflow velocities ranging from $-1000$ to $-6200$ km~s$^{-1}$. These properties distinguish them from the slower warm absorbers (WAs) flowing at only hundreds of km s$^{-1}$. Interestingly, these ionized obscurers have been observed exclusively in AGNs that also have WAs.

Here we detect an ionized obscurer in Mrk 817, and for the first time, we show that it is kinematically linked to UV broad absorption lines (Fig.~\ref{fig:UV}). Moreover, we do not detect a WA making it the first example of an obscurer without a WA. Perhaps shielding of high-energy photons by the obscurer prevents the launching of WAs at larger radii or perhaps WAs have different covering factors than the obscurers.

The fact that the UV and X-ray absorption components are kinematically linked, and that the column density of the obscurer traces the equivalent width of UV lines  on longer timescales of months (\citealt{2023arXiv230212896P}) indicates that the low ionization wind found in the UV and the higher ionization wind found in the X-rays are spatially coincident. This spatial coincidence can be physically manifested in different ways. UV absorption may arise from weakly ionized gas clumps embedded within a more ionized, diffuse X-ray obscuring medium as suggested by \citet{1995ApJ...447..512K} and supported by the variability of the UV and X-ray covering fraction of the absorbing gas (\citealt{2023arXiv230212896P}). Alternatively, different ionization zones may result from outflow stratification, with the most ionized X-ray layers nearest the black hole and the UV layers further away, consistent with the UV obscurer's slightly lower velocity compared to the X-ray obscurer (Fig.~\ref{fig:UV}) and the impact of the order of the photoionization components on the outflow characteristics in our spectral models (see Section 4.3 for more details).
Based on the results of these models, we can place constraints on the location and duty cycle (Section \ref{sec:radius}), energetics (\ref{sec:kinLum}), and thermodynamic properties (4.3) of the outflow.

\subsection{Constraints on the Launching Radius and Energetics}\label{sec:radius}

The ionization parameter of the absorbing plasma 
\begin{equation}
\xi=\frac{L_{\text {ion }}}{n r^2}
\label{eq:ioniz}
\end{equation}
is a function of the radial distance of the absorber from the ionizing source ($r$), the electron number density ($n$), and the ionizing luminosity $L_{\mathrm{ion }}$ calculated over $1$-$1000$ Ryd ($13.6$ eV to $13.6$ keV). While we can measure the ionizing luminosity, the distance and the density of the absorber are degenerate and cannot normally be directly determined from spectral modelling.

Using constraints on the ionization parameter, column density and velocity of the obscurer, we can make fairly crude, order-of-magnitude estimates for the location of the outflow. All {\sc pion} components have relatively similar velocities, so we make the assumption that these components trace gas at a similar location. Taking the arithmetic mean of the parameters of the time-averaged model fits in Table~\ref{tab:params} and then summing up the resulting column densities, we adopt a total column density of $ N_{\mathrm{H}}^{\mathrm{tot}} \simeq 6.6 \times 10^{22}$ cm$^{-2}$, an ionization parameter of $\log \xi \simeq 3.1$ i.e $\xi \simeq 1300$  cm erg~s$^{-1}$, an outflow velocity of $v$ $\simeq-5200$ km~s$^{-1}$, and a $1$-$1000$ Ryd ionizing luminosity of $L_{\text {ion}} \simeq 8.1 \times 10^{44}$ erg~s$^{-1}$.

Assuming that the thickness ($\Delta r$) of the absorbing layer is not larger than its distance from the X-ray source (i.e $\Delta r/r < 1$) and using $N_{\mathrm{H}}\sim n C_v \Delta r \sim C_v (L/\xi) (\Delta r / r^2)$ , where $C_v < 1$ is the volume filling factor, we can set an upper limit for the distance of the obscurer from the black hole of 
$$r_{\max }=\frac{L_{\mathrm{ion}}}{N_{\mathrm{H}} \xi} \approx 1.6 \times 10^{6} R_g \simeq 9.5 \times 10^{18} \: \mathrm{cm} \simeq  3 \: \mathrm{pc} $$ (e.g, \citealt{2012ApJ...753...75C,2013MNRAS.430.1102T, 2021A&A...654A..32S}). 
 This estimate assumes that the absorber is confined to a small radial range at a well defined distance from the central source. If we also assume that the outflow velocity is equal to the local escape velocity, then the distance of the obscurer from the central source is estimated as
   $$ r = 2 \frac{c^2}{v^2} R_g \approx 6600 \: R_g \simeq 3.8 \times 10^{16} \mathrm{cm} \simeq 0.01 ~ \mathrm{pc},$$
where we assume a black hole mass of $\log (M_\mathrm{BH}/M_\odot) = 7.59_{-0.07}^{+0.06}$ (\citealt{2018ApJ...864..146B}). We note that the winds need not have the local escape speed. \textcolor{black}{This estimate is an upper limit on the distance only if Mrk 817’s outflow is a failed wind, implying the total outflow velocity is less than the escape velocity and hence the line of sight velocity is also lower than the escape velocity. If the outflow is successful, the total outflow velocity exceeds the escape velocity, but the line of sight velocity is not necessarily lower or greater than the escape velocity.}

\textcolor{black}{Additionally, we can use the time variability of the third photoionization component during our observation to improve our distance estimate. If we assume that the observed variability is due to transverse motion across the line of sight, then we can estimate the transverse velocity of this component given the size of the X-ray source. Lewin et al.(submitted) used the 2021~Apr XMM-Newton observation of Mrk 817 to perform frequency-resolved X-ray reverberation mapping, which revealed short soft X-ray lags ($\sim150$~s). These lags correspond to a coronal size of approximately~$\sim~10~R_g$, consistent with reverberation modeling in other AGN, suggesting a compact corona (e.g, \citealt{2016MNRAS.462..511K,2010ApJ...709..278D,2016AN....337..356C,2022ApJ...939..109L}).}

Thus, assuming $R_c \approx 10 R_g$ for the coronal size, and given that we observe variability over ($\delta t \approx 60$~ks), the implied transverse velocity is at least 
$$ \frac{R_c}{\delta t}  \approx 9500 \: \mathrm{km}\, \mathrm{s}^{-1}. $$
For simplicity, we assume a Keplerian velocity for the absorbing clumps, then the \textcolor{black}{part of the outflow modeled by {\sc pion} 3} should be located at a distance of at most
$$r \approx 1000 R_g \simeq 5.7 \times 10^{15} \mathrm{cm} \simeq 0.002 ~\mathrm{pc} $$ from the black hole. \textcolor{black}{This estimate of the distance places the third photoionization component near the inner broad line region, consistent with the distance to the obscurer inferred by the reverberation time delays between the UV-emitting disk and the broad C~IV emission line (\citealt{2023arXiv230211587H}). It also aligns with the estimates in Paper~1 for the distance of the obscurer based on UV photoionization modeling with {\sc CLOUDY} (\citealt{2023RMxAA..59..327C}). Thus, the distance this outflow crosses during $60$~ks if it was indeed located at the $1000 R_g$ inferred from the UV analyses is comparable to the X-ray corona size, in line with the possibility that the observed variability during our observation is due to transverse motion across the line of sight. Thus, while this variability-based distance estimate, in our analysis, is inferred for {\sc pion} 3, it is plausible that all the X-ray photoionization components are co-spatial. This is supported by their similar velocities ($4000 - 6000$~km~s$^{-1}$) and the agreement with the independent distance measurements from the UV, given the kinematic consistency between the X-ray and UV outflow that we established in section \ref{sec:UV}.}

Next, we can determine the corresponding particle density of the wind as $n=L_{ion}/(r^2\xi) \simeq 2\times 10^{10}$~cm$^{-3}$ which is consistent with the CLOUDY density estimate in Paper 1 made using broad UV absorption troughs. 
\newline
\newline
Finally, we can make a rough estimate of the time it takes for the outflow to leave the disk and reach our line of sight
$$t_{\rm flight} =  \frac{\text{distance travelled}}{\text{velocity of gas}} \sim \frac{\tan(90\degree - 30\degree) \ 1000 \ \mathrm{R_g}}{5200 \ \mathrm{km~s^{-1}}}$$ yields
$$t_{\rm flight} \sim 200 \ \mathrm{days},$$ 

where we have assumed an inclination of $30 \degree$ (Paper~1 constrains the inclination to be $< 40 \degree$ and, independently, \citet{2021ApJ...911L..12M} constrains it to $\sim 20 \degree$). \textcolor{black}{We assume that the wind launched at the 1000~$R_{\mathrm{g}}$ distance estimate (making the flight time a lower limit if it was launched below 1000~$R_{\mathrm{g}}$)}, and assume that the vertical velocity of the outflow to be similar to the line of sight velocity of $\sim 5200$ km~s$^{-1}$.  
If it takes $\sim 200$ days for material to reach our line of sight, then we only sample a small part of the wind streamline, far from the launching site during our brief $\sim$ 120~ks XMM-Newton observation. Any variations at the launching point (e.g. higher mass outflow rate due to increased luminosity) that could result in differences in the column density or velocity structure likely take months to reach our line of sight. The only exception is the ionization parameter, which responds quickly to changes in the ionizing continuum if the wind density is sufficiently high. Interestingly, 200~days is roughly the timescale on which we observe large, order of magnitude changes in the X-ray column density of the obscurer (\citealt{2023ApJ...947....2P}). \textcolor{black}{Additionally, if we instead use the escape radius of $6600~R_{\mathrm{g}}$, we obtain a flight time of $\sim 1500$~days which is too slow to reproduce the observed variability in the EW of the UV absorption troughs and changes in the X-ray obscuration observed throughout the campaign.} 
\newenvironment{colorthisnote}{\par\color{brown}}{\par}

\begin{figure*}[!tbp]
\centering
\includegraphics[width=\textwidth]{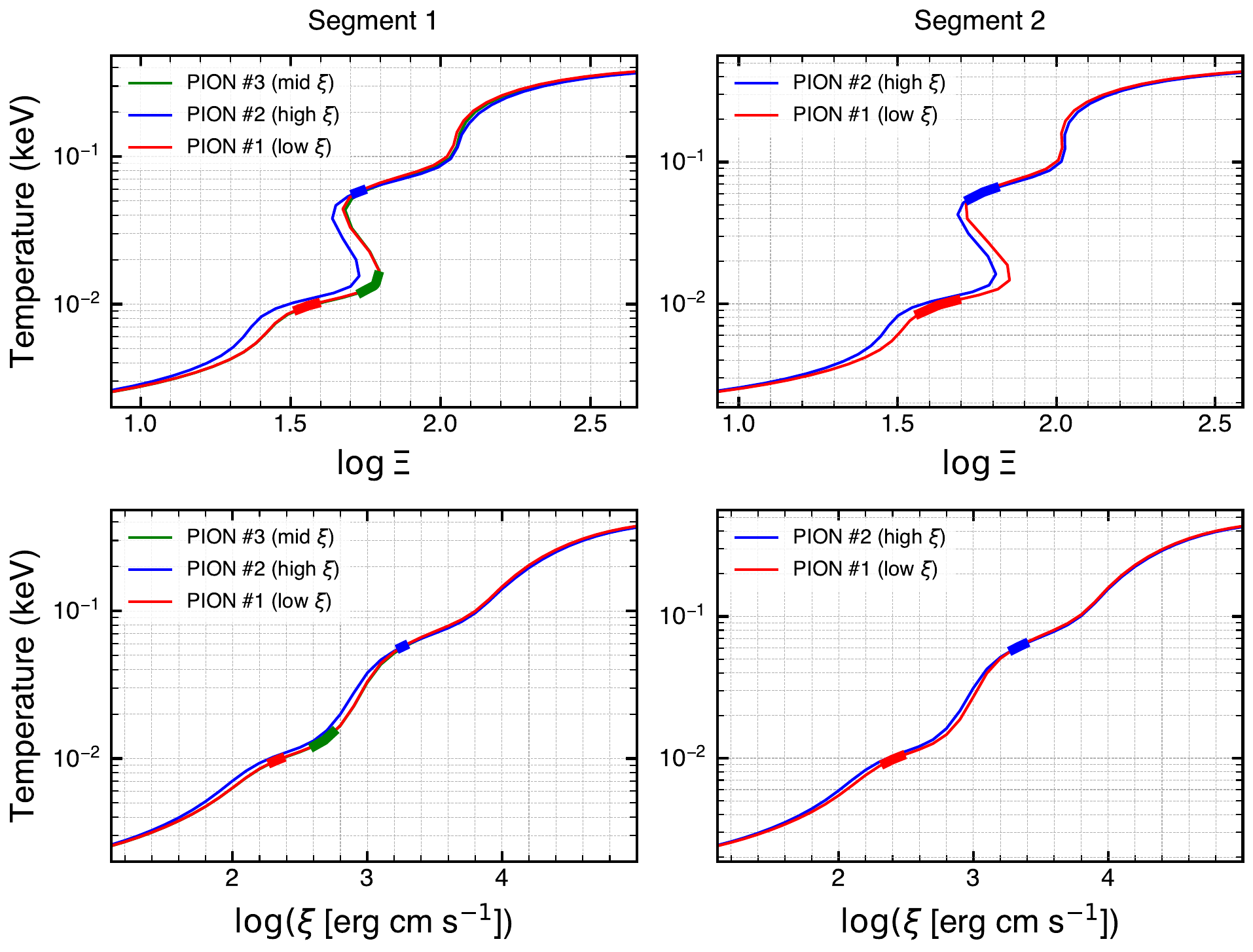}
\caption{Thermal stability curves (S-curves) showing the equilibrium electron temperature of the gas as a function of the ionization parameter ($\xi$) in the bottom panels and as a function of the pressure-~form of the ionization parameter ($\Xi$) in the top panels. The S-curves are calculated using the SEDs seen by each of the photoionization components during the first half of the observation (Segment 1, left column) and the second half (Segment 2, right column). The thick strips show the positions of the derived photoionization components (including the fitting uncertainties) from the best-fitting model of the spectra.}
\label{fig:Scurve}
\end{figure*}

\subsection{Estimates of the mass outflow rate and kinetic luminosity}\label{sec:kinLum}

The continuity equation $\dot{M} = \Omega r^2 v \rho$
relates the mass loss rate to the radius, velocity and
solid angle $\Omega$ of the outflow.  The outflow mass density
$\rho = C_v m_p n$ depends on the electron number density $n$
and the filling factor of the absorbing outflow $C_v$.  The
electron number density is also related to the total column density $N_{\mathrm{H}}^{\mathrm{tot}}= n C_v \Delta r$ where $\Delta r$ is the thickness of the absorbing layer.
Combining these and assuming that $\Delta r < r$ where $r$ is the distance from the X-ray source to the outflow, the
lower limit on the mass loss rate is

\begin{equation}
       { \dot{M} \over \Omega } > r m_p N_{\mathrm{H}}^{\mathrm{tot}} v
      \simeq 0.005~M_\odot yr^{-1}
      \label{equ: Mdot}
\end{equation}

similar to Equation (1.1) of \citet{2013IAUS..290...45K}, where we have used $r \sim 1000~R_g$, $v \simeq -5200$~km~s$^{-1}$, and
$N_{\mathrm{H}}^{\mathrm{tot}} \simeq 6.6 \times 10^{22}$~cm$^{-2}$. This
mass loss rate then implies a minimum kinetic
luminosity of

$$
    \frac{L_{\mathrm{kin}}}{\Omega} \gtrsim 4.2 \times 10^{40} \text{ erg~s$^{-1}$ $\Omega^{-1}$} \sim 0.003 \% ~L_{\mathrm{bol}}\Omega^{-1}
$$

where $L_{\mathrm{bol}} \sim 1.4 \times 10^{45} ~\text{erg~s$^{-1}$}$ is calculated from our unobscured SED model ($10^{-3}$ to $10^5$ keV).

To obtain an upper limit, we follow \citet{2005A&A...431..111B}, where we can use Eq.~\ref{eq:ioniz} to determine $n$, leading to 
\begin{equation}
\label{eq:blustin}
\frac{\dot{M}}{C_v~\Omega} \sim \frac{m_p L_{\rm ion}  v }{\xi}.
\end{equation}

We use the same values for $\xi$, $L_{\text {ion}}$, and $v$ as before. 

$$ \frac{L_{\mathrm{kin}}}{\Omega C_v} \lesssim 7.4 \times 10^{43} \text{erg~s$^{-1}$ $\Omega^{-1} C_v^{-1}$} \sim 5 \% ~L_{\mathrm{bol}}\Omega^{-1} C_v^{-1}$$ 
The value of $C_v$ is unknown, we can obtain an upper limit on the kinetic luminosity by assuming $C_v =1$. For a reasonable value of $\frac{\Omega}{4\pi} \sim 0.25 $ we obtain a range of kinetic luminosity of $0.01 \% ~L_{\mathrm{bol}}$ to $16 \% ~L_{\mathrm{bol}}$ which is too broad to determine the significance of feedback. 
If we instead use $C_v \sim 8\%$ \textcolor{black}{ as the upper limit on the volume covering factor derived by \citealt{2005A&A...431..111B} from many similar AGN}, using Equation \ref{eq:blustin}, we obtain a kinetic luminosity of $1.3 \% ~L_{\mathrm{bol}}$ which is above the $0.5\%$ threshold needed for significant feedback (\citealt{2010MNRAS.401....7H}). This holds for any $C_v > 3.1\%$ in our calculation using $\frac{\Omega}{4\pi}~\sim~0.25$. Thus, based on this exercise, it is a possibility that the outflow plays a role in driving feedback on large scales in Mrk 817. Assessing the outflow's true impact on the broader galactic environment necessitates estimating the total kinetic energy released over a significant part of the AGN's lifetime, this kinetic luminosity estimate is from our one day observation, which might not hold true over longer timescales. Long term monitoring campaigns, such as AGN STORM 2, will help us better constrain the duty cycle.

\subsection{Thermal state of the photoionized gas}\label{sec:thermal}

The {\sc spex} photoionization modeling estimates the thermal and ionization balance based on the SED of the ionizing source, the elemental abundances of the ionized plasma (\citealt{2016A&A...596A..65M}), and assume photoionization equilibrium. Under this assumption, the temperature of the gas is a unique function of its ionization parameter $\xi$.
If we assume pressure balance between the radiation field and the gas then the pressure form of the ionization parameter is (\citealt{1981ApJ...249..422K}): 
$$ \Xi=\frac{F}{n_{\mathrm{H}}c k T} =\frac{L_{\mathrm{ion}}}{4 \pi r^2 n_{\mathrm{H}} c k T}=\frac{\xi}{4 \pi c k T} \approx 19225 \frac{\xi}{T}$$ 
where $F=L_{\mathrm{ion}}/(4 \pi r^2)$ is the flux of the ionizing source between $1$-$1000$ Ryd and $k$ is the Boltzmann constant.
The electron temperature T shown as a function of $\Xi$ is often referred to as the S-curve or the thermal stability curve. In regions of the curve with a positive gradient, a small increase in temperature will increase the cooling, and the photoionized gas is thermally stable. Regions with a negative gradient are thermally unstable since a small perturbation in temperature will instead grow.

Fig.~\ref{fig:Scurve} shows the S-curves for Segments 1 and 2. Each photoionization model sees a slightly different SED, which yields a slightly different S-curve. The photoionization components are arranged so that the highest ionization component sees the unobscured SED, and the lowest ionization component sees an obscured SED. While this is a reasonable assumption, the actual order of the photoionization components is not known, and the shape of the SED is sensitive to the order of the components.

Fig.~\ref{fig:Scurve} shows that {\sc pions} 1 and 2 lie in thermally stable regions for both segments (red and blue stripes). On the other hand, {\sc pion} 3 of Segment 1 falls near the thermal instability region of the S-curve. This could explain its variability over the timescale of the observation, as slight variations in the conditions of the gas can lead to significant fluctuations in the observed properties of this wind component.

The top two panels of Fig. \ref{fig:Scurve} show that the unstable regions of the S curves, i.e where the gradient is negative, for {\sc pion} 3 (green) and {\sc pions} 1 (red) span a slightly wider range of $\Xi$ compared to the blue S curve ({\sc pion} 2). This is not surprising since {\sc pions} 1 and 3 see a more obscured SED. A higher level of absorption or obscuration of the ionizing radiation can affect the ionization balance and temperature structure of the gas and increase the likelihood of variability in the observed properties of the gas (\citealt{2015A&A...575A..22M}).

\section{Summary and Conclusions}\label{sec:summa}

As part of the multi-wavelength monitoring program AGN STORM 2 of the Seyfert 1 galaxy Mrk~817, we conducted an X-ray campaign consisting of four XMM-Newton observations, all of which were accompanied by simultaneous NuSTAR observations. 
The focus of this paper is the 2021~April observation. This was the highest flux observation from our campaign, due to a reduced obscuration and an increased intrinsic luminosity. The greater flux allowed us to probe the properties of the obscurer with high resolution spectroscopy. We summarize the results as follows:
\begin{itemize}
    \item Modeling the RGS spectra requires three photoionization components with varying column densities, ionization parameters, and velocities (4000-6000 km~s$^{-1}$), which reveals the multiphase nature of the outflow. 

    \item Comparing the kinematics between the X-ray absorption line profiles from the XMM-Newton/RGS spectrum and UV absorption line profiles from the HST/COS spectrum, taken a day prior, provides kinematic evidence for the first UV counterpart detected for an X-ray ionized obscurer. 
    \item \textcolor{black}{The lower velocity component of the X-ray absorber shows variability on short ($\sim60$ ks) timescales, aligning with the overall obscurer's location of $\sim 1000~R_{\mathrm{g}}$ inferred from independent UV measurements, and consistent with an accretion disk wind.}

    \item Analysis of the thermal state of the photoionized gas indicates that the variable photoionization component is positioned close to the thermally unstable region on the S-curve.

    \item The time it takes for material to travel from the accretion disk into our line of sight is roughly $200$~days, consistent with the large-scale column density changes seen in long timescale X-ray monitoring with NICER (\citealt{2023ApJ...947....2P}).

    \item The kinematic luminosity, $L_{\mathrm{kin}} \sim 1.3 \% ~L_{\mathrm{bol}}$,  is sufficiently high for AGN feedback, assuming a solid angle of $\pi$ and a volume filling factor of 8\% as often observed for AGN.

\end{itemize}

\section{Acknowledgements}

FZ and EK acknowledge support from NASA grant 80NSSC22K0570. PK acknowledges support from NASA through the NASA Hubble Fellowship grant HST-HF2-51534.001-A awarded by the Space Telescope Science Institute, which is operated by the Association of Universities for Research in Astronomy, Incorporated, under NASA contract NAS5-26555. EB is grateful for the hospitality of MKI, where this collaborative work took place. Research at UC Irvine was supported by NSF grant AST-1907290. HL acknowledges a Daphne Jackson Fellowship sponsored by the Science and Technology Facilities Council (STFC), UK. MCB gratefully acknowledges support from the NSF through grant AST-2009230. D.I., A.B.K and L.C.P. acknowledge funding provided by the University of Belgrade--Faculty of Mathematics (contract No. 451-03-47/2023-01/200104) and Astronomical Observatory Belgrade (contract No. 451-03-47/2023-01/200002) through the grants by the Ministry of Science, Technological Development and Innovation of the Republic of Serbia. D.I. acknowledges the support of the Alexander von Humboldt Foundation. A.B.K. and L.C.P. thank the support by Chinese Academy of Sciences President’s International Fellowship Initiative (PIFI) for visiting scientist. CSK is supported by NSF grant AST-2307385. M.V. gratefully acknowledges financial support from the Independent Research Fund Denmark via grant number DFF 8021-00130.

\newpage 
\appendix
\begin{figure}[H] 
\begin{center}
    \includegraphics[width=\textwidth]{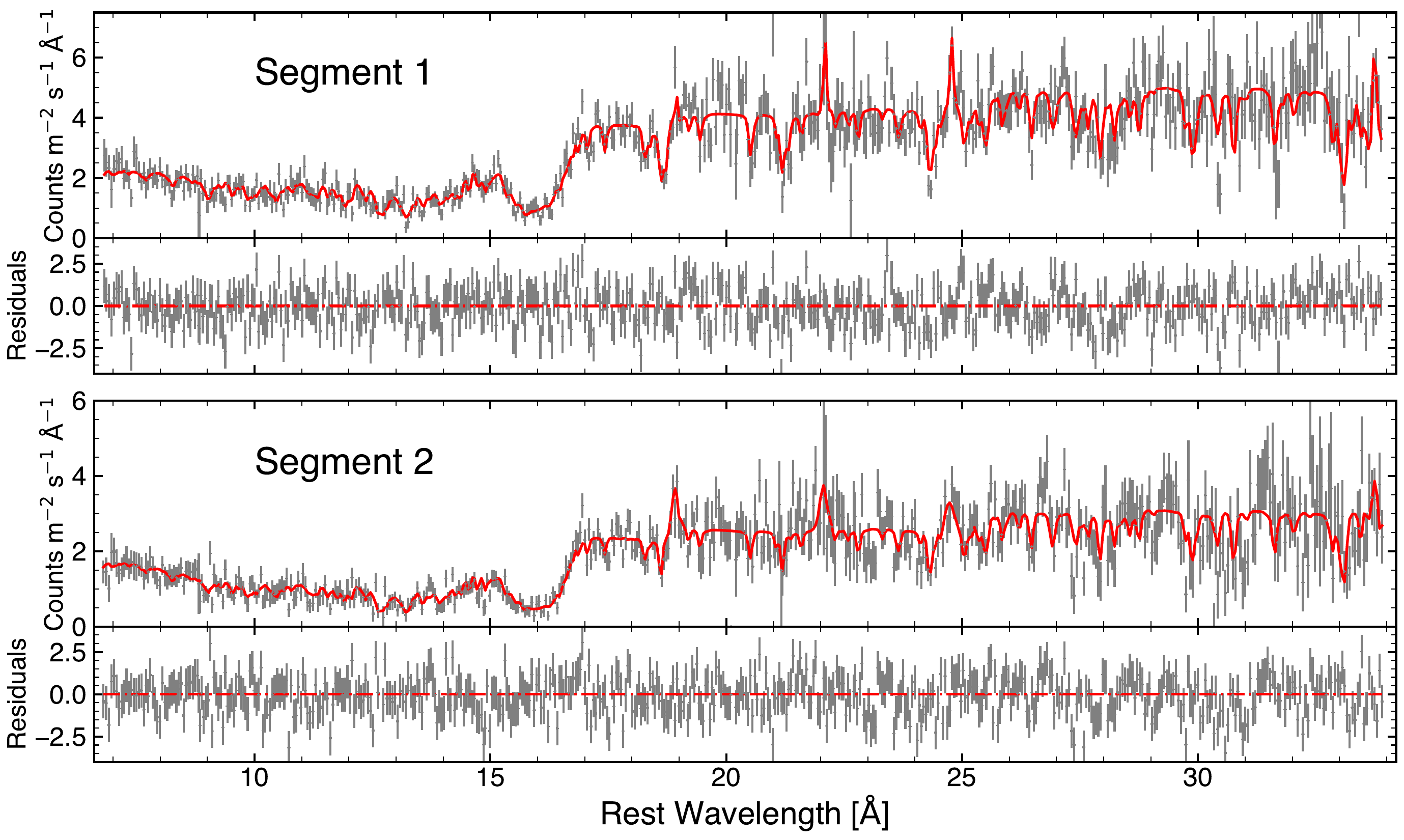}
    \caption{The RGS spectra for Segment 1 (top panel) and Segment 2 (bottom panel) with the best fitting model overlaid in red. The residuals are calculated as “(model-data)/error”. 
    }
\label{fig:RGS_segments}
\end{center}
\end{figure}

\label{appendix}

\bibliographystyle{apj}
\bibliography{ref.bib}

\end{document}